\documentclass[12pt]{article} 
\usepackage{jheppub}

\usepackage{amsmath,amsfonts,amsbsy,amssymb,array,accents,dsfont}
\usepackage{enumerate}
\usepackage[shortlabels]{enumitem}
\usepackage{graphicx,verbatim} 
\usepackage{caption} 
\usepackage{subcaption} 
\usepackage{mathrsfs}
\usepackage{upgreek}
\usepackage{bm}
\usepackage{slashed}
\usepackage{ytableau} % per i simboli Young tableaux

\usepackage{enumitem}
\setlist[itemize,1]{label=\textbullet}  % Forza il bullet normale

\usepackage{float}
\usepackage{hyperref}
\usepackage{overpic}
\usepackage{xcolor}
\allowdisplaybreaks

\title{%\LARGE 
The hyperplane string, RCFTs, and the swampland}

\author{Guglielmo Lockhart and Luca Novelli}

\affiliation{
Bethe Center for Theoretical Physics, Universit\"at Bonn, D-53115, Germany
}

\emailAdd{glockhar@uni-bonn.de}
\emailAdd{lnovelli@uni-bonn.de}

\abstract{ 
Six dimensional $\mathcal{N}=(1,0)$ supergravity features BPS strings whose properties encode highly nontrivial information about the parent 6d theory. We focus on a distinguished set of theories whose string charge lattice is one-dimensional. In geometric theories, the generator of the lattice arises from a D3 brane wrapping the hyperplane class in $\mathbb{P}^2$. This hyperplane string is expected to remain stable even when one ventures beyond the geometric regime where it becomes challenging to verify which candidate 6d theories belong to the swampland. We identify five 6d models which from the perspective of the hyperplane string deviate the most from being geometric. For these theories we are able to provide an exact description of the left-moving sector of the hyperplane string worldsheet theory in terms of a rational conformal field theory and provide evidence for their consistency. In one instance, using RCFT methods we are able to determine the elliptic genus and find that in the unflavored limit it matches with the elliptic genus of geometric models. We argue that the non-geometric model is connected to geometric ones via a sequence of Higgsing transitions. These results lead us to formulate a proposal relating the quantum corrected moduli space of the hyperplane string CFT with a region of the landscape of 6d $\mathcal{N}=(1,0)$ quantum gravity.
}

\begin{document}

\maketitle

\section{Introduction}

There has been a growing realization in recent years that  the worldvolume theory of extended probes in quantum gravity provides an important tool for exploring the landscape of consistent theories and discriminating between effective field theories that admit a consistent coupling to gravity and those which fall within the swampland
\cite{KimShiuVafa2019,Lee:2019skh,Kim:2019ths,Katz:2020ewz,Angelantonj:2020pyr,Hamada:2021bbz,Tarazi:2021duw,Cheng:2021zjh,Cvetic:2021vsw,Bedroya:2021fbu,Long:2021jlv,Dierigl:2022zll,Hayashi:2023hqa,Kim:2024tdh,KimVafaXu2024}. In the context of six-dimensional theories, unitarity of the worldsheet theory of certain classes of 1-brane probes is a powerful guiding light. Among other things, it shines a light on the possible gauge symmetries that can be realized, due to the fact that bulk gauge symmetries make an appearance on the worldsheet under the guise of left-moving current algebras. This in particular puts a hard bound on the rank of the gauge groups that can be attained, and it also puts constraints on the 6d anomaly coefficients, which control the amount by which the current algebras contribute to the left-moving central charge. For theories with $T\geq 1$, significant mileage can be obtained by considering the properties of a class of strings known as \emph{H-strings} \cite{KimVafaXu2024}.

For 6d theories that can be realized within F-theory, a vast geometric toolbox is available that allows for stringent tests of the swampland conjectures \cite{Grimm:2018ohb,
Blumenhagen:2018nts,
Lee:2018urn,
Grimm:2018cpv,
Joshi:2019nzi,
Font:2019cxq,
Lee:2019xtm,
Erkinger:2019umg,
Lee:2019wij,
Grimm:2019ixq,
Baume:2019sry,
Bastian:2020egp,
Bastian:2021eom,
Palti:2021ubp,
Klawer:2021ltm,
Alvarez-Garcia:2021mzv,
Alvarez-Garcia:2021pxo,
Grimm:2022sbl,
Rudelius:2023odg,
Alvarez-Garcia:2023gdd,
Hayashi:2023hqa,
Alvarez-Garcia:2023qqj,
Kim:2024tdh,
KimVafaXu2024,
Aoufia:2025ppe}.
Geometric considerations also place demands on supergravity theories such as the Kodaira condition \cite{Kumar:2010ru} and the existence of a neutral universal hypermultiplet which, in hindsight, have been understood not to be fundamental requirements of quantum gravity theories, but rather the consequence of restricting to a specific region of the landscape. Indeed several examples of non-geometric constructions within string theory have been identified which give rise to 6d theories that violate the geometric constraints. This includes models that can be constructed using orbifolds \cite{Kachru:1995wm,Gkountoumis:2023fym}, asymmetric Gepner models \cite{Israel:2013wwa}, and free fermions \cite{Dolivet:2007sz}.

This raises the question of how the stable brane probes, whose properties are well understood in geometric settings, behave as one ventures away from the geometric realm and what they can teach us in non-geometric settings. This idea has been recently employed \cite{Hayashi:2023hqa} to study BPS strings of a non-geometric model with $SU(2)$ gauge group and matter that cannot be realized geometrically, and has been used to verify that this model satisfies several of the swampland conjectures.

One class of theories that merits attention are those whose string charge lattice is one dimensional. In a geometric setting this class is engineered by elliptic threefolds fibered over $\mathbb{P}^2$, and the charge lattice is generated by the hyperplane $[L]$. This class covers almost all of the known cases that do not possess H-strings in their spectrum\footnote{An exceptional candidate has also been proposed with $T=9$ \cite{KimVafaXu2024}.}. A large database of candidate 6d supergravity theories with $T=0$ has been compiled by the authors of \cite{HamadaLoges2023} which provides a rich set of examples to examine. Absent an H-string, it is natural to ask what the hyperplane string associated to the hyperplane $[L]$ of $\mathbb{P}^2$ may have to teach us. We denote this string as $L$-string for short. In the large volume limit, it is known that, provided that the elliptic fiber does not degenerate along $[L]$, the $L$-string admits a description in terms of a heterotic NLSM of central charges $(c_L,c_R) = (32,12)$ whose target space is a fibration $K : T^4\to\mathbb{P}^2$ \cite{HaghighatMurthyVafaVandoren2015}. One basic expectation that proves to be significant is that in the large volume limit the chiral algebra component of the string worldsheet CFT which descends from the 6d gauge symmetry should be realized by the heterotic bundle on the left-moving side, and therefore it should satisfy a bound
\begin{equation}
c_L^{gauge}\leq c_L - \text{dim} K = 24.
\label{eq:bound}
\end{equation}
This gives us a worldsheet criterion to single out models which cannot possibly be geometric. A scan of the database of \cite{HamadaLoges2023} reveals that many of the entries correspond to models that violate the bound \eqref{eq:bound}; this includes a number of models which also do not satisfy the Kodaira condition or possess no neutral hypermultiplets.

 One way that has been proposed to connect geometric models to non-geometric ones is to take a limit where the size of the F-theory base becomes stringy and a transition leads to the freezing of the universal hypermultiplet \cite{BaykaraHamadaTaraziVafa2023, Baykara:2024vss}, analogously to the quantum versions of conifold transitions that have been conjectured to occur in four dimensions \cite{Kachru:1995wm}. On the other hand it is also known that quantum corrections can drastically alter the moduli space of heterotic sigma models \cite{Melnikov:2019tpl}, and it is natural to wonder if the breakdown of the NLSM description in these models may be correlated to the breakdown of a geometric description within F-theory.

This analogy suggests to focus on theories that maximally violate the bound \eqref{eq:bound} in order to better understand the behaviour of $L$-strings in the non-geometric regime. In doing this, we make a startling observation: these theories are also the ones for which the $L$-string CFT undergoes a dramatic simplification! In particular we find a set of five theories within the list of \cite{HamadaLoges2023} for which
\begin{equation}
31<c_L^{gauge}<32.
\end{equation}
These models are listed in Table \ref{tab:clg31intro} and are all models with no uncharged matter that violate the Kodaira condition. We are not aware of any existing ways to test whether these theories are in the swampland, but the CFT perspective gives us exactly this. Namely, the corresponding 2d CFTs can only be unitary if $c_L^{res} = 32 - c_L^{gauge}$ coincides with the central charge of a unitary Virasoro minimal model. This turns out to be precisely the case.
\begin{table}[t!]
\centering
\begin{tabular}{c|c|c|c}
\textbf{Current algebra} & \textbf{Charged Matter} $\mathcal{H}$ & \textbf{$c^{gauge}_L$} & \textbf{$c^{res}_L$} \\
\hline
$A_{7, 8}$  & $\mathbf{336}$ & $63/2$ & 1/2 \\
$A_{5,6} \oplus D_{4, 6}$  & $(\mathbf{56},\mathbf{1}) +  (\frac{1}{2}\mathbf{20}, \mathbf{28})$ &  $63/2$ & 1/2\\
$A_{3,6} \oplus A_{3,6} \oplus E_{7,2}$  & $(\mathbf{10},\mathbf{10}, \mathbf{1}) + (\mathbf{1}, \mathbf{6},\frac{1}{2}\mathbf{56}) + (\mathbf{6}, \mathbf{1},\frac{1}{2}\mathbf{56})$ & $313/10$ & $7/10$\\
$A_{3,10} \oplus A_{11,2}$ & $(\mathbf{35},\mathbf{1}) +  (\mathbf{6}, \mathbf{66})$ & $218/7$ & 6/7\\
$E_{6, 8}$ & $\mathbf{351}'$ & $156/5$ &  4/5 \\
\end{tabular}
\caption{List of 6d anomaly-free models with $c_L^{gauge} \ge 31$.}
\label{tab:clg31intro}
\end{table}

The implication is that the $L$-string in these maximally non-geometric models is governed by an RCFT, whose left-moving sector is completely fixed by simple unitarity considerations. Determining the right moving sector is a fascinating problem which we hope to report on soon \cite{wip}. Rationality of the CFT in fact proves to be sufficient to determine the elliptic genus for the model of Table \ref{tab:clg31intro} with $E_{6}$ current algebra at level 8, which couples to the $c_L^{res}=4/5$ three-state Potts model. The model appears to possess an unexpected symmetry under the $S_3$ Dynkin diagram automorphism group of affine $E_6$, and it would be very interesting to understand if similar phenomena occur for the other RCFTs. Moreover we find that its elliptic genus, turning off $E_6$ fugacities, matches with the elliptic genus of the model with trivial gauge group corresponding to F-theory on the  $X_{18}(1,1,1,6,9)$ threefold \cite{HuangKatzKlemm2015}, and we identify a set of candidate theories with intermediate gauge groups that potentially connect the two via a sequence of Higgsing transitions:
\begin{equation}
E_6 \xrightarrow{} F_4 \xrightarrow{} B_4 \xrightarrow{} D_4 \xrightarrow{} B_3 \xrightarrow{} G_2 \xrightarrow{} A_2
\xrightarrow{} A_1 \xrightarrow{} \emptyset.
\label{tab:higgsingchain6d}
\end{equation}
More details about these models are presented in Table \ref{tab:higgsing2} below.

\begin{figure}[h!]
\centering
\includegraphics[width=\textwidth]{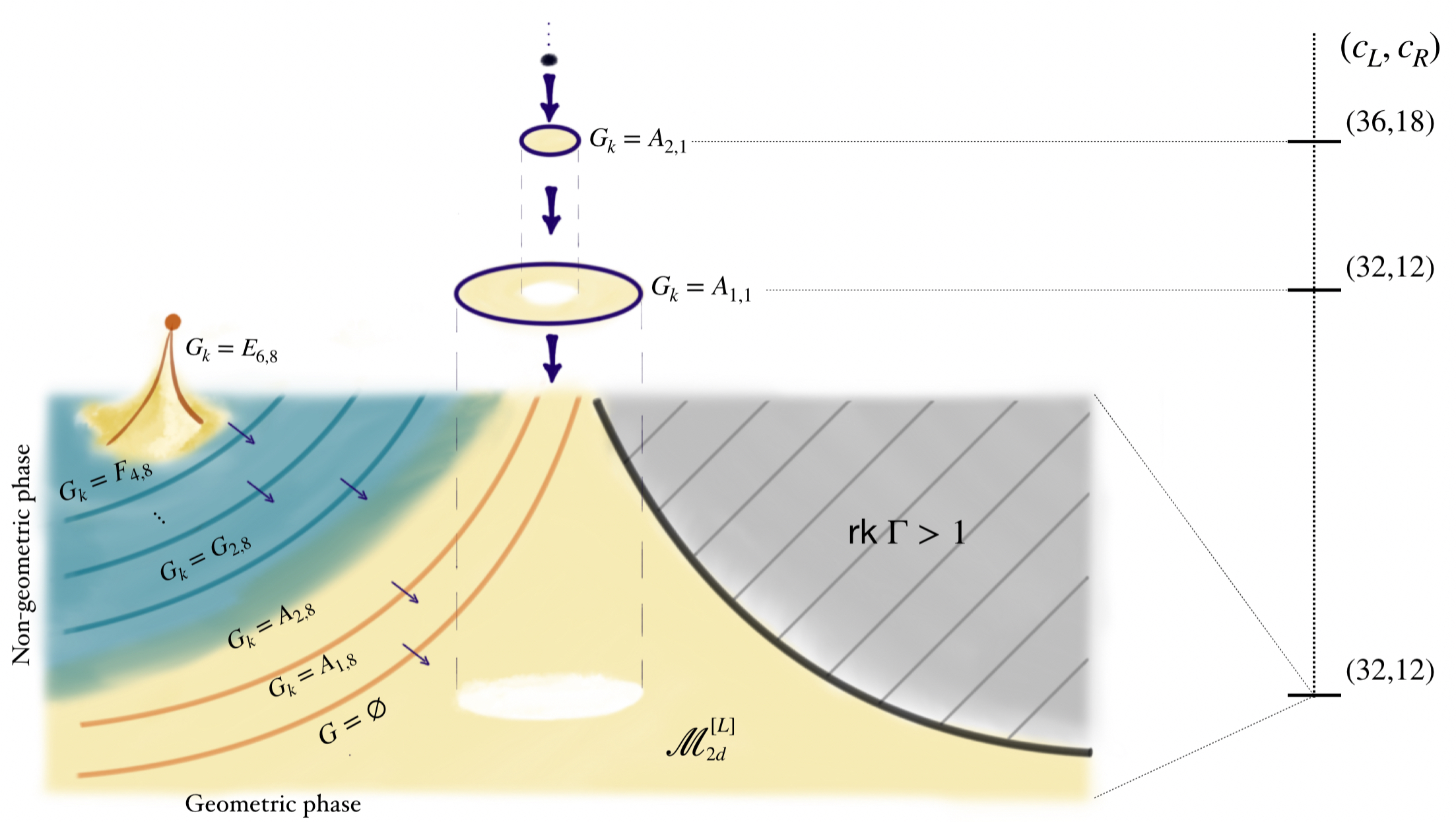}
\caption{Cartoon of the quantum corrected moduli space $\mathcal{M}^{[L]}_{2d}$ of the hyperplane string CFT. Higgsing transitions of the parent 6d theory are depicted by arrows and are of two kinds: the thick vertical arrows start from 2d worldsheet theories corresponding to 6d theories with nontrivial gauge group and $b_i = 1$; in this case, the 2d moduli space also includes instanton branches in addition to the geometric branch, resulting in potentially higher central charges. This sequence of Higgsings ultimately terminates on $\mathcal{M}^{[L]}_{2d}$. The thin arrows correspond to Higgsing of gauge factors for which $b_i\neq 1$. In this picture we have portrayed a series of transitions that start from the theory with $G=E_6$ and matter in the $\mathbf{351'}$ dimensional representation, and terminate on the fully Higgsed theory with trivial gauge group and 273 neutral hypermultiplets. The $E_6$ theory is portrayed as an island where the string worldsheet theory is described by an RCFT. This Higgsing chain involves passing through a number of non-geometric models (portrayed here as the sea) as well as geometric models (portrayed as the sand). Branches of the moduli space where the string becomes a composite object (for $\text{rk } \Gamma > 1$) are grayed out. We have displayed only one chain of Higgsing transitions here, but there also exist other regions (see Table \ref{tab:clg31intro}) with other islands connected to the geometric mainland by chains of Higgsing transitions.}
\label{fig:landscape}
\end{figure}

Taking stock of the preceding comments, we have ventured into a seemingly very non-geometric region of the landscape and found that this leads to a much simpler description for the $L$-string. This is reminiscent of the simplification that occurs in other settings within string theory in going away from the geometric regime, famous examples including 
Gepner points in the moduli space of Calabi-Yau manifolds \cite{Gepner:1987qi,Eguchi:1988vra,Nahm:1999ps} and,  more generally, the Landau-Ginzburg/CY correspondence \cite{Martinec:2001hh,Greene:1988ut}. This suggests the following natural picture: in the large volume region, the moduli of the target space of the NLSM for the $L$-string are closely related to the moduli of the F-theory compactification. As one moves away from this limit, quantum corrections become important in both pictures, and the NLSM description breaks down. Nonetheless, we propose that the 2d $\mathcal{N}=(0,4)$ CFT on the string remains consistent and its quantum corrected moduli space ${\mathcal{M}}^{[L]}_{2d}$ (depicted schematically in Figure \ref{fig:landscape}) maps surjectively onto the region of the landscape ${\mathcal{M}}^{[L]}_{QG}$ corresponding to theories with one-dimensional string-charge lattice and gauge anomaly coefficients $b_i\neq 1$:
\begin{equation}
{\mathcal{M}}_{2d}^{[L]} \twoheadrightarrow  {\mathcal{M}}^{[L]}_{QG}.
\end{equation}
This explains how the $L$-string is able to access information about non-geometric models and suggests that its worldsheet theory might be holding further lessons about the 6d landscape in store. Our perspective in this regard is that the $L$-string, being a stable BPS object, is a useful tool to study the landscape of consistent 6d quantum gravity theories without tensors even if a UV complete description is not yet well understood. This includes not just studying specific models and testing whether they are in the swampland, but also studying global aspects such as the existence of connected paths (which involve Higgsing transitions) between regions of the landscape, which can be tested by verifying the matching of supersymmetry-protected quantities (in this case, the elliptic genus).\\

The remainder of this paper is organized as follows:
in Section \ref{subseq:AnomalyFree} we briefly review necessary conditions for a 6d $\mathcal{N}=(1,0)$ supergravity theories with $T = 0$ tensor multiplets to be free of gauge and gravitational anomalies; in Section \ref{subsec:Geomvsnongeom} we review the geometric F-theory setup, we discuss criteria a 6d theory must satisfy to admit a geometric description and give examples of candidate non-geometric theories; in Section \ref{subsec:Higgs} we discuss Higgsing transitions between (both non-geometric and geometric) theories.
Section \ref{sec:[L]string} is devoted to explaining the role of hyperplane strings in 6d $\mathcal{N}=(1,0)$ quantum gravity. In Section \ref{eq:hyper} we review the description of $L$-strings in geometric models and formulate our proposal relating the quantum corrected moduli space ${\mathcal{M}}^{[L]}_{2d}$ of the $L$-string CFT to the region $\mathcal{M}^{[L]}_{QG}$ in the 6d landscape. In Section \ref{sec:ng} we discuss our worldsheet criterion for identifying 6d models which are necessarily non-geometric and identify a set of extremal examples for which the left-moving sector of the worldsheet is captured by a RCFT. We also propose a chain of Higgsings that connect these theories to geometric models. Section \ref{sec:ell} discusses general properties of the elliptic genus of $L$-strings and its relation to the 6d massless spectrum.
Section \ref{sec:E6string} is devoted to a detailed study of the model with $E_6$ gauge symmetry, for which the left-moving sector of the string CFT is given by the tensor product of the $E_{6}$ affine Lie algebra at level 8 and of the three-state Potts model; in Section \ref{sec:computation} we determine the elliptic genus using techniques inspired by rational conformal field theory and the theory of vector valued modular forms, while in Section \ref{eq:ellcomm} we comment on the unexpected symmetries of the model and the interesting behavior of the string spectrum under modular transformations. We present our conclusions in Section \ref{sec:disc} where we also discuss a number of avenues of future research. The appendices contain technical details: Section \ref{app:mf} contains our basic definitions for modular forms; Section \ref{app:rcft} contains details about the three-state Potts model and the affine $E_6$ Lie algebra relevant to Section \ref{sec:E6string}, and \ref{app:cons} contains the details of our Ansätze for the elliptic genus of the $E_6$ model.
\newpage

\section{6d quantum gravity with $\text{rk}(\Gamma)=1$: geometric and non-geometric models}
\label{sec:geom}
In Section \ref{subseq:AnomalyFree}, we review anomaly cancelation conditions for 6d $\mathcal{N}=(1,0)$ supergravity theories. In Section \ref{subsec:Geomvsnongeom}, after summarizing the geometric construction of certain 6d models within F-theory, we summarize known necessary conditions for a theory to admit a geometric realization and we discuss some examples of non-geometric anomaly-free theories.
Finally, in Section \ref{subsec:Higgs}, we discuss a chain of 6d theories including both non-geometrically and geometrically realizable ones, which we will argue are related by Higgsing transitions in Sections \ref{sec:[L]string} and \ref{sec:E6string}.
\subsection{Review of anomaly-free conditions}
\label{subseq:AnomalyFree}
In this section we briefly review anomaly-free consistency conditions of 6d $\mathcal{N}=(1,0)$ supergravity theories: our presentation follows that of \cite{Kumar:2010ru, HamadaLoges2023, KimVafaXu2024}. We focus on the non-Abelian case, whose field content is composed of $V$ vector multiplets transforming under a choice of gauge group $G=\prod_{i=1}^n G_i $, a number $T$ of tensor multiplets, a gravity multiplet and $H_{\text{tot}}=H_0+H_c$ hypermultiplets, where $H_0$ and $H_c$ denote respectively the number of neutral and charged hypers. The matter corresponding to $H_c$ is charged under a (possibly reducible) representation of $G$, which we denote as $\mathcal{H}$.\\
The supersymmetric massless spectrum is organized as in Table \ref{tab:masslessspectra}, where $+$ or $-$ indicate the chirality for spinors and the self-duality or antiself-duality of the field strength for a given 2-form $B^I_{\mu\nu}$, with $I= 0,1,..., T$. The objects that are charged under these $T+1$ 2-forms are strings whose charges live in a lattice $\Gamma \subset \mathbb{R}^{1,T}$. \\
The chiral index \cite{Lee:2018urn} of the massless spectrum of 6d supergravity theories is determined in terms of the field content as:
\begin{equation}
n_0 = 2 (V+3+T-H_{\text{tot}})
\end{equation}
where the contribution of $+3$ comes from the gravitino. \\
The Green-Schwarz-Sagnotti mechanism \cite{Green:1984,Green:1985,Sagnotti:1992} leads to a set of constraints that are required for a theory to be anomaly-free, which are determined by imposing that the 1-loop anomaly polynomial $I_8$ factorizes as follows:
\begin{equation}
I_8=\frac{1}{2} \Omega_{\alpha \beta} X_4^\alpha X_4^\beta, \quad X_4^\alpha=\frac{1}{2} a^\alpha \operatorname{tr} R^2+\sum_{i=1}^n b_i^\alpha \frac{2}{\lambda_i} \operatorname{tr} F_i^2
\label{eq:anomalypolynomialI8factorization}
\end{equation}
where $\Omega_{\alpha \beta}$ is the metric on $\Gamma$, $R$ is the curvature 2-form, $F_i$ the gauge field strength, $a^\alpha$ and $b_i^\alpha$ correspond to the gravitational and gauge anomaly coefficients, respectively. This factorization can be equivalently expressed as:
\begin{equation}
    \begin{aligned}
& H_{\text{tot}}-V+29 T=273, \quad a \cdot a=9-T \\
& B_{\mathbf{a d j}}^i=\sum_{\mathbf{r}} n_{\mathbf{r}}^i B_{\mathbf{r}}^i, \quad a \cdot b_i=-\frac{\lambda_i}{6}\left(\sum_{\mathbf{r}} n_{\mathbf{r}}^i A_{\mathbf{r}}^i-A_{\mathbf{a d j}}^i\right), \\
& b_i \cdot b_i=\frac{\lambda_i^2}{3}\left(\sum_{\mathbf{r}} n_{\mathbf{r}}^i C_{\mathbf{r}}^i-C_{\mathbf{a d j}}^i\right), \quad b_i \cdot b_j= \lambda_i \lambda_j \sum_{\mathbf{r}, \mathbf{s}} n_{\mathbf{r}, \mathbf{s}}^{i j} A_{\mathbf{r}}^i A_{\mathbf{s}}^j \quad i \neq j,
\label{eq:anomalyfreeconstraintsNONABELIAN}
\end{aligned}
\end{equation}
where $n_{\mathbf{r}}^i$ and $n_{\mathbf{r}, \mathbf{s}}^{i j}$ are the number of matter fields in the representation $\mathbf{r}$ of $G_i$ and $\mathbf{r} \otimes \mathbf{s}$ of $G_i \times G_j$, respectively. Moreover, $a$ and $b_i$ generate an integral sub-lattice of the string charge lattice $\Gamma$. The $\lambda_i$ are normalization constants depending on the specific simple factors $G_i$ and can be found for example in Table 1 of \cite{Kumar:2010ru} while the coefficients $A_{\mathbf{r}}$, $B_{\mathbf{r}}$ and $C_{\mathbf{r}}$ are defined as $\operatorname{tr}_{\mathbf{r}} F^2=A_{\mathbf{r}} \operatorname{tr} F^2$, $\operatorname{tr}_{\mathbf{r}} F^4=B_{\mathbf{r}} \operatorname{tr} F^4+C_{\mathbf{r}}\left(\operatorname{tr} F^2\right)^2$. \\
In the following, we will only deal with the $T=0$ case, where the anomaly coefficients $a$ and $b_i$ are $n+1$ integers. We will use the notation $\mathcal{T}_{G,\mathcal{H}, H_0}$ to denote a 6d supergravity theory without tensors and $\mathcal{I}(G,\mathcal{H}, H_0)$ to refer to the chiral index, which is given by
\begin{equation}
    \mathcal{I}(G,\mathcal{H}, H_0) = 2V-2H_{\text{tot}}+6.
\end{equation}

\begin{table}[t]
\centering
\begin{tabular}{c|c}
\textbf{Multiplet} & \textbf{Field Content} \\
\hline
Gravity & \( \left(g_{\mu \nu}, \psi_\mu^{+}, B_{\mu \nu}^{+} \right) \) \\
Tensor  & \( \left(\phi, \chi^{-}, B_{\mu \nu}^{-} \right) \) \\
Vector  & \( \left(A_\mu, \lambda^{+} \right) \) \\
Hyper   & \( \left(4 \varphi, \psi^{-} \right) \) \\
\end{tabular}
\caption{Massless spectrum of a $6d$ $\mathcal{N} = (1,0)$ supergravity theory.}
\label{tab:masslessspectra}
\end{table}

\subsection{Geometric vs. non-geometric}
\label{subsec:Geomvsnongeom}
In this section we briefly review the F-theory geometric realization of 6d theories and review known constraints that follow from requiring geometric constructions. We follow this by providing some interesting examples of non-geometric candidate supergravity theories. The geometric set-up consists of a Calabi-Yau Threefold $X$ that admits an elliptic fibration with section over a Kähler base space $B$ of complex dimension two which, for theories without tensor multiplets ($T=0$), is simply given by ${\mathbb{P}}^2$. \\
In this setting, the 6d non-abelian gauge group $G$ is determined by the codimension-one singularities of the elliptic fibration that are specified by the discriminant locus in the base
\begin{equation}
    \Delta = 4f^3 + 27g^2,
\end{equation}
while the codimension-two singularities determine the localized-matter content. \\
In the abelian gauge group case, the situation is more complicated since the Mordell-Weyl group of the fibration needs to be taken into account, as discussed in \cite{MorrisonPark2012}.
In order to realize an elliptic fibration equipped with a Calabi-Yau structure, $f$, $g$ and $\Delta$ must be sections of specific line bundles over the base \cite{CompactificationFtheory1, CompactificationFtheory2}
\begin{equation}
    f \in -4 K_B,\qquad g \in -6 K_B,\qquad \Delta \in -12 K_B
\label{eq: consistencyconditionforellipticfibration}
\end{equation}
where, for $T=0$, the canonical divisor $K_B$ is a multiple of the hyperplane class $L$ of the base ${\mathbb{P}}^2$, namely $K_B = - 3L$. Any divisor on ${\mathbb{P}}^2$ is a multiple of the hyperplane class since $H^{1,1}({\mathbb{P}}^2) \cong H^{2}({\mathbb{P}}^2)$ is one-dimensional. Focusing on a gauge group written as a direct product of non-abelian simple factors $G=\prod_{i=1}^n G_i
$, the discrimant locus is described by the divisor $\Delta$ in the base and can be decomposed into a sum
\begin{equation}
    -12 K_B = 36 L = \Delta = \sum_{i=1}^n \nu_i \Sigma_i + Y
    \label{eq: KodairaconditionGeometry}
\end{equation}
where the $\Sigma_i$ are irreducible effective divisors in ${\mathbb{P}}^2$ giving rise to the gauge group factors $G_i$, the coefficients $\nu_i$ are the order of vanishing of $\Delta$ along the $\Sigma_i$ and the residual divisor $Y$ is a sum of effective divisors associated to non-contractible curves that satisfy $Y \cdot \Sigma_i \ge 0 $.

Equation \eqref{eq: KodairaconditionGeometry} is known as the Kodaira condition and is a necessary condition for a 6d theory without tensors to be realized geometrically in F-theory. 
The Kodaira condition can be rephrased straightforwardly in terms of the anomaly coefficients. Indeed, as shown in \cite{Kumar2009, Kumar:2010ru}, one can map information concerning anomaly-free conditions of the $6d$ theory to topological data of the base of the elliptic fibration, as briefly summarized in Table~\ref{tab:bottomupvsgeometry}. We use the notation $\Gamma_{\text{BPS}}$ for the BPS cone, which is a subset of the string charge lattice $\Gamma$ generated by primitive BPS strings, namely those with charges that cannot be written as a non-negative linear combination of charges of other BPS strings \cite{KimVafaXu2024}. \\
Hence, Equation \eqref{eq: KodairaconditionGeometry} can be written as 
\begin{equation}    
    -12 a = \sum_{i=1}^n \nu_{i} b_{i} + Y
\end{equation}
or, equivalently, as
\begin{equation}    
    - j \cdot ( 12 a + \sum_{i=1}^n \nu_{i} b_{i}) \ge 0
\label{eq: kodairaconditionEFTlevel}
\end{equation}
since the tensor branch scalars $j$ satisfy $j \cdot Y \ge 0$.  \\
Additional constraints follow from the positivity of the 6d gravitational and gauge kinetic terms, which imply that
\begin{equation}
j \cdot a <0\qquad \text{and} \qquad j \cdot b_i>0
\end{equation}
respectively. \\

It has been recently recognized that the Kodaira condition, while necessary to ensure a geometric F-theory construction, is not required for consistency of a 6d supergravity theory. Indeed it was realized in \cite{BaykaraHamadaTaraziVafa2023} that asymmetric orbifold constructions lead to consistent models which violate the Kodaira constraint and are thus intrinsically non-geometric.  
Furthermore, in any known F-theory realization, the number of neutral hypermultiplets is always greater than zero, as one can naively deduce from
\begin{equation}
    H_0 = h^{2,1}(X) + 1.
\end{equation} 
In fact, there always exists at least one neutral hypermultiplet, known as the \emph{universal hypermultiplet}, which originates from the volume modulus of the geometry after compactification. The absence of neutral hypermultiplets is therefore a marker that signals the absence of a geometric description \footnote{Although this is not a necessary feature of non-geometric theories: for example the $SU(2)$ model with one triple-symmetric and 84 fundamental hypermultiplets considered in \cite{Hayashi:2023hqa} is non-geometric but possesses 75 neutral hypers.}. Examples of theories with $H_0 = 0$ have nonetheless been discovered within the string theory landscape by other methods \cite{Kachru:1995wm,Dolivet:2007sz,Israel:2013wwa,Gkountoumis:2023fym}. Remarkably, it was recently realized \cite{BaykaraHamadaTaraziVafa2023, CVSG25} that certain 6d models that can be constructed as asymmetric orbifolds \cite{Narain:1986qm, Narain:1990mw, Kachru:1995wm} are in fact connected to the landscape of geometric models. Namely, one can consider a limit where the base $B$ of the elliptic threefold shrinks to Planckian size, which takes us outside of the Type IIB supergravity regime as the volume of the base scales as the inverse of the 6d gravitational coupling:
\begin{equation}
G_6 \sim \frac{1}{Vol(B)}.
\label{eq:g6}
\end{equation}

 In this limit, stringy effects become important and quantum transitions can lead to intrinsically non-geometric models. For instance, asymmetric orbifolds can be used to construct theories with $H_0 = 0$, such that, after Higgsing all charged multiplets, one obtains a geometric compactification on a Calabi–Yau threefold. The expectation is that these regions are connected by a transition in the stringy regime of the compactification, in which the volume modulus itself can become part of the charged hypermultiplet sector~\cite{BaykaraHamadaTaraziVafa2023}.

\begin{table}[H]
\centering
\begin{tabular}{c|c}
\textbf{Bottom-up (charge lattice)} & \textbf{Top-down (F-theory)} \\
\hline
$\Gamma$ & $H_2(B, \mathbb{Z})$  \\
$\Gamma_{BPS}$ & Effective divisors in $H_2(B, \mathbb{Z})$  \\
$a=-3$ & $K_{B}=-3L$ \\
$b_i$ & $\Sigma_i = b_i L$ \\
$j$ & $J = j L$ \\
\end{tabular}
\caption{Dictionary between string charge lattice and F-theory data in geometric $6d$ $\mathcal{N}=(1,0)$ supergravity theories with $T=0$. Here, $J$ is the Kähler class of the base space B.}
\label{tab:bottomupvsgeometry}
\end{table}
By performing a systematic search for 6d supergravity theories that are anomaly free, the authors of \cite{HamadaLoges2023} have compiled a large database of candidate supergravity theories \cite{database}. The database includes theories with gauge groups $G_i\neq U(1),A_1,A_2$ and matter charged simultaneously under at most two factors. While this is not a complete list, it nonetheless provides a large set of interesting examples of candidate supergravity theories of which many are necessarily non-geometric. As a candidate non-geometric model, let us consider the following $T=0$ theory which will be the test case for our ideas below:
\begin{equation}
G=E_{6}, \quad \mathcal{H} = 1 \times \mathbf{351}', \quad b_i = 8,
\label{eq: theoryE6,8}
\end{equation}
which is anomaly-free as it satisfies the constraints \eqref{eq:anomalyfreeconstraintsNONABELIAN}. 
However, this model cannot be realized in F-theory for a few reasons. First, the Kodaira condition \eqref{eq: kodairaconditionEFTlevel} is not satisfied as $\nu_i = 8$ for $E_6$. 
Moreover, the theory contains only charged hypermultiplets, i.e., $H_0 = 0$, which is another reason why it is non-geometric. \\

We can consider other $T=0$ theories similar to the one above. For example,
\begin{equation}
G=F_{4}, \quad \mathcal{H} = 1 \times \mathbf{324}, \quad b_i = 8
\label{eq: theoryF4,8}
\end{equation}
is anomaly-free, but it still lacks a geometric realization. In this case, $H_0 \ge 1$, so one cannot draw definitive conclusions about its (non-)geometric nature based solely on $H_0$. However, the Kodaira condition is still violated, confirming that this is a non-geometric model. In general, one might expect non-geometric theories to be difficult to study. However, we will show that this is not the case. In fact, in Sections \ref{sec:[L]string} and \ref{sec:E6string} we will find that the most non-geometric 6d quantum gravity theories are precisely the ones in which the worldsheet theory on the BPS strings simplifies as it is captured by a RCFT.

\subsection{Higgsing down to geometric models}
\label{subsec:Higgs}
It is natural to consider possible ways to Higgs the gauge group of $\mathcal{T}_{E_6,\mathbf{351'},0}$. For example, one can consider a chain of Higgsing transitions that progressively reduce the gauge group as follows:
\begin{equation}
E_6 \xrightarrow{} F_4 \xrightarrow{} B_4 \xrightarrow{} D_4 \xrightarrow{} B_3 \xrightarrow{} G_2 \xrightarrow{} A_2
\xrightarrow{} A_1 \xrightarrow{} \emptyset.
\label{tab:higgsingchain6d}
\end{equation}
The corresponding matter content is in Table \ref{tab:higgsingchainmodels}.

Let us consider for instance the first transition in \eqref{tab:higgsingchain6d}, namely $E_{6} \xrightarrow{}F_{4}$. In this case, the $6d$ matter content of $E_6$ sits in the representation $\mathbf{351'}$. The relevant $E_6$ representations branch into $F_4$ irreps as follows:
\begin{align}
    \mathbf{351'}_{E_6} &= \mathbf{324}_{F_4} \oplus \mathbf{26}_{F_4} \oplus \mathbf{1}_{F_4} \label{eq:351decomp} \\
    \mathbf{78}_{E_6}   &= \mathbf{52}_{F_4} \oplus \mathbf{26}_{F_4}. \label{eq:adjointdecomp}
\end{align}
Hence, it is possible to give mass to 26 gauge bosons by absorbing the same number of scalars via Higgsing,  leading to the $F_4$ model with matter content $ 1 \times \mathbf{324}_{F_4}$ presented in Table \ref{tab:higgsingchainmodels}. Similarly, one can progressively Higgs the gauge group to lower entries in the table all the way to the fully ungauged theory. All gauge groups appearing in the table have $b_i = 8$. 
\begin{table}[H]
\centering
\begin{tabular}{c|c|c}
{$G$} & $H_0$ &$\mathcal{H}$ \\
\hline\hline
$E_6$ & $0$ & $\mathbf{351}'$\\
%\hline
$F_4$ & $1$ & $\mathbf{324}$ \\
%\hline
$B_4$ & $2$ & $\mathbf{9} + \mathbf{44} + \mathbf{126} + \mathbf{128}$\\
%\hline
$D_4$ & $4$ & $\mathbf{8}_v + \mathbf{8}_s + \mathbf{8}_c + \mathbf{35}_v + \mathbf{35}_s + \mathbf{35}_c + \mathbf{56}_v + \mathbf{56}_s + \mathbf{56}_c$ \\
%\hline
$B_3$ & $6$ & $ \mathbf{7} + 4 \times \mathbf{8} + \mathbf{21} +  \mathbf{27} + 3 \times  \mathbf{35} + 2 \times  \mathbf{48}$ \\
%\hline
$G_2$ & $13$ & $10 \times \mathbf{7} + 3 \times \mathbf{14} + 6 \times  \mathbf{27}$\\
%\hline 
$A_2$ & $29$ &  $36 \times \mathbf{{3}} + 9 \times \mathbf{8} + 12 \times \mathbf{6}$
\\
%\hline
$A_1$ & $85$ & $21 \times  \mathbf{3} + 64 \times  \mathbf{2}$ \\
%\hline
$\emptyset$ & $273$ & No charged matter \\
%\hline
\end{tabular}
\caption{Matter content of the 6d theories presented in the Higgsing chain \eqref{tab:higgsingchain6d}.}
\label{tab:higgsingchainmodels}
\end{table}

The entries with $G=E_6\supset\dots\supset G_2$ all appear in the database \cite{database} and are all non-geometric as they violate the Kodaira condition. On the other hand, the $A_2$ model can be realized geometrically in F-theory as discussed in \cite{Klevers:2017aku}. Similarly the geometry has been explicitly constructed for the $A_1$ model \cite{, Hayashi:2023hqa} as well as for the case with trivial gauge group \cite{Candelas1993mirrorI, Candelas1994mirrorII, Hosono1995hypersurfaces, Hosono1995complete, HuangKatzKlemm2015, HaghighatMurthyVafaVandoren2015}.

\section{The hyperplane string and the 6d $\mathcal{N}=(1,0)$ landscape}
\label{sec:[L]string}
We now turn our focus to the $L$-strings and propose a worldsheet approach to explore the landscape of 6d $\mathcal{N}=(1,0)$ quantum gravity theories and novel tests for the consistency of anomaly-free models. In Section \ref{eq:hyper} we introduce the hyperplane string in the context of geometric compactification on F-theory fibrations over $\mathbb{P}^2$, we discuss features of the heterotic sigma model that describes it, and propose a relation between its moduli space and the landscape of 6d quantum gravity that extends beyond the geometric regime. In Section \ref{sec:ng} we find a set of anomaly-free theories that are necessarily non-geometric, and give an explicit description of the left-moving sector of the hyperplane string as a rational CFT. We also discuss Higgsing transitions that connect these models with geometric ones. Finally in Section \ref{sec:ell} we introduce the elliptic genus of the hyperplane string and discuss its relation with the massless spectrum of 6d theories.

\subsection{The hyperplane string}
\label{eq:hyper}
For the moment, let us stay in the realm of 6d $\mathcal{N}=(1,0)$ supergravity theories which admit a geometric realization as F-theory compactified on an elliptic threefold $X$ at large volume. We have restricted our attention to theories with $T= 0$, so the internal geometry is given by an elliptic fibration
\begin{equation}
X \to \mathbb{P}^2.
\end{equation}

This setting is especially simple as the second homology of the base manifold $\mathbb{P}^2$ is one-dimensional:
\begin{equation}
H_2(\mathbb{P}^2,\mathbb{Z}) = \mathbb{Z},
\end{equation}
with the hyperplane class $[L]$, which is dual to the K\"ahler two-form $J$ of $\mathbb{P}^2$, serving as generator. A D3 brane that wraps the hyperplane gives rise to a primitive BPS string in 6d which couples to the anti-symmetric two-form field in the gravity multiplet and generates the entire lattice. We denote this string as the hyperplane string, or $L$-string for short. 

Going beyond geometric models, the question of whether this BPS string is always part of the spectrum has been investigated in \cite{KimVafaXu2024}. Strong arguments have been put forward why this should be the case for any consistent supergravity theory with $T=0$ by considering its compactification to 5d, where one finds BPS particles which must originate as states of a BPS string in the original 6d theory. This is consistent with expectations based on the completeness of spectrum hypothesis \cite{Polchinski:2003bq,Banks:1997zs,Harlow:2018tng}, applied to 6d supergravity \cite{KimShiuVafa2019,Tarazi:2021duw} theories. Additionally, there is an expectation that the BPS string remains a primitive object in the BPS cone as the 6d theory undergoes Higgsing transitions \cite{Tarazi:2021duw}, which do not affect the string charge lattice and in particular do not lead to the appearance of additional strings that may destabilize a primitive BPS string. These considerations are valid beyond the geometric realm, and it has been shown that the supersymmetric spectrum of the string as captured by the elliptic genus can be tracked across Higgsing transitions between geometric and non-geometric theories. The supersymmetric spectrum of the string, as captured by the elliptic genus, in particular allows for precision tests of the Weak Gravity Conjecture \cite{Hayashi:2023hqa} along the lines of \cite{Lee:2018urn,Lee:2018spm}. In Section \ref{sec:E6string} we are able to determine the elliptic genus for the $L$-string in the 6d theory $\mathcal{T}_{E_6,\mathbf{351'},0}$ introduced in Section \ref{subsec:Geomvsnongeom}, which is deep in the non-geometric region, and find that in its unrefined limit it reduces to the elliptic genus of the geometric theory $\mathcal{T}_{\emptyset,\mathbf{0},273}$. This gives further evidence that one can trust the $L$-string even after moving far into the non-geometric regime.\\

A meaningful question is how the region in the landscape of 6d quantum gravity theories where the $L$-string is a primitive object, which we will denote as $\mathcal{M}^{[L]}_{QG}$, is realized from the perspective of the worldsheet of the string. We propose a natural interpretation that neatly explains our results concerning the $\mathcal{T}_{E_6,\mathbf{351'},0}$ theory and opens up a new approach to explore this region of the landscape.

It is again useful to start by considering geometric models, where the worldsheet theory on the string can be obtained by topologically twisted compactification \cite{Bershadsky:1995vm} of the worldvolume theory of wrapped D3 branes \cite{HaghighatMurthyVafaVandoren2015}. The 6d theory may have nontrivial gauge symmetry arising from the degeneration of the elliptic fiber along codimension-one loci in $B$ \footnote{ For simplicity of exposition we restrict our discussion here to non-abelian symmetries.} given by
\begin{equation}
\Sigma_i = b_i [L],
\end{equation}
in which case one must make a distinction based on the values of $b_i$. Namely, either all $b_i \neq 1$, in which case the string has a purely geometric moduli space, or some $b_i=1$ in which case the string carries instanton charge with respect to the corresponding gauge group \footnote{ See in particular \cite{Hayashi:2023hqa} for examples where either type of scenario is realized.} and its moduli space includes a branch corresponding to an instanton moduli space. While the 2d gravitational anomaly, which is proportional to $c_L-c_R $, is the same in both cases, the left- and right-moving central charges differ and are minimal in the case of all $b_i\neq 1$. This was pointed out initially in the case of SCFT strings in \cite{DelZotto:2017mee,DelZotto:2018tcj} and holds true just as well in supergravity theories \cite{Hayashi:2023hqa}. One can explicitly realize Higgsing transitions as complex structure deformations of the elliptic threefold which remove the components of the discriminant with $b_i=1$ from the discriminant locus. This can be viewed as triggering an RG flow for the 2d theory on the string to a 2d $\mathcal{N}=(0,4)$ CFT with lower central charges. While it is certainly very interesting to consider cases where some $b_i=1$, in what follows we will focus exclusively on the region within where all $b_i \neq 1$. We denote this region as ${\mathcal{M}}^{[L]}_{2d}$.

On general grounds one expects the worldsheet theory on the string to be described in this region by a $\mathcal{N}= (0,4)$ 2d nonlinear sigma model, whose central charges and anomaly coefficients have been computed geometrically \cite{HaghighatMurthyVafaVandoren2015} and coincide with the values obtained \cite{KimShiuVafa2019} by anomaly inflow \cite{Kim:2016foj, Shimizu:2016lbw} without relying on geometric considerations. For a generic supergravity BPS string associated to an element $Q\in \Gamma_{BPS}$ of the string charge lattice, the central charges are given by \footnote{ We always remove the center of mass factor of central charges $(c_L,c_R)=(4,6)$ and contributes a decoupled free hypermultiplet.}
\begin{align}
c_L &=3 Q \cdot Q-9a \cdot Q+2,
\\
c_R&=3 Q \cdot Q-3 a \cdot Q,
\\
k_i&=Q \cdot b_i,
\end{align}
for the hyperplane string these reduce to
\begin{align}
c_L=32,\qquad c_R=12,\qquad k_i=b_i\neq 1.
\label{eq:ccharges}
\end{align}

In fact, more can be said about the sigma model for the $L$-string. In the large volume limit, the target space of the sigma model can be viewed as a fibration over a $\mathbb{P}^2$ which parametrizes the moduli space of deformations of the hyperplane curve inside the base of the threefold. The fiber includes a $T^4$ which is common to the left- and right-moving sector of the CFT. Besides this, the fiber includes an additional 24-dimensional Narain torus giving rise to left-moving degrees of freedom \cite{HaghighatMurthyVafaVandoren2015}. The $T^4$ fibration over $\mathbb{P}^2$ must give rise to a hyperkahler manifold as a consequence of $\mathcal{N}=4$ supersymmetry, while the Narain torus gives rise to a bundle on the left-moving side that satisfies the usual properties required for consistency of heterotic sigma models \cite{Witten:1985xe,Freed:1986zx,Distler:1986wm,Dai:1994kq,Melnikov:2019tpl}.  \\

An analogous picture holds when one considers the heterotic string in compactifying F-theory on a Hirzebruch surface $\mathbb{F}_n = \mathbb{P}^1_f\to\mathbb{P}^1_b$. There, the heterotic string arises from a D3 brane wrapped on the rational fiber $\mathbb{P}^1_f$, whose moduli space of deformations coincides with the base $\mathbb{P}^1_b$ \cite{Witten:1985xe,Freed:1986zx,Distler:1986wm,Dai:1994kq,Melnikov:2019tpl}. Twisted compactification of the D3 brane worldvolume theory gives rise to an additional $T^2$ fibered over $\mathbb{P}^1$, which gets reinterpreted as the K3 manifold on which the heterotic string is compactified. Finally, the bundle of left movers in this case has rank 16 and gives rise to the heterotic bundle. \\

What is the significance, then, of the target space of the NLSM for the $L$-string? Crucially, its base $\mathbb{P}^2$ originates from the geometric $\mathbb{P}^2$ of the F-theory compactification, so we expect  its K\"ahler parameter to be related to the strength of the gravitational coupling of the 6d supergravity theory. Moving away from the large volume limit we expect the NLSM description of the $L$-string to break down due to quantum corrections, which goes hand in hand with the breakdown of the geometric description within F-theory as well as of the supergravity approximation. We propose that the quantum corrected moduli space $\mathcal{M}_{2d}^{[L]}$ of the NLSM maps surjectively onto the region $\mathcal{M}_{QG}^{[L]}$ of the landscape:
\begin{equation}
\mathcal{M}^{[L]}_{2d}
\twoheadrightarrow
\mathcal{M}^{[L]}_{QG}.
\end{equation}
In particular, $\mathcal{M}_{2d}^{[L]}$ includes regions that do not admit a geometric description within F-theory. We remark that, as long as all $b_i \neq 1$, Higgsing transitions by which the gauge symmetry of the 6d theory is reduced do not change the central charges; from the perspective of the string, these transitions correspond to moving between different branches on the moduli space of the theory. A schematic picture of our proposal is summarized in Figure \ref{fig:landscape} in the Introduction.\\

In the next section we will discuss the fate of the hyperplane string in non-geometric regions of ${\mathcal{M}}_{QG}^{[L]}$ and find that we can reach points on the landscape where the worldsheet theory of the string drastically simplifies and admits a RCFT description.

\subsection{The non-geometric region and RCFT}
\label{sec:ng}

As we have seen in the previous section, from the point of view of the string we expect that in a theory that admits a geometric realization the gauge symmetry should be captured by a current algebra arising from the 24-dimensional Narain lattice, which therefore should contribute
\begin{equation}
    c^{gauge}_L = \sum_i \frac{k_i \cdot \text{dim} G_i}{k_i + h^{\vee}_{G_i}} \leq 24
\label{eq:unitarityconditionBPSstring}
\end{equation}
to the central charge, where $k_i=b_i$ is the level of the current algebra supported on component $\Sigma_i$ of the discriminant. In fact, considering that the bundle of left-movers should be fibered nontrivially over $\mathbb{P}^2$, one would expect $c^{gauge}_L $ to be strictly less than 24. Nevertheless, unitarity of the worldsheet theory of the string by itself only requires $c_L^{gauge} \le 32$, and this leads us to propose a worldsheet criterion to identify theories which are necessarily non-geometric:
\begin{equation}
c_L^{gauge} > 24 \quad \rightarrow \quad \text{No geometric realization is possible.}
\label{eq:centralchargeandgeometryFtheory..}
\end{equation}
We expect that a correct description of these models requires taking into account stringy effects that would not be visible in a large volume compactification in F-theory. We stress that this is a different criterion from the Kodaira constraint or the absence of neutral hypermultiplets, and it would be interesting to understand if it can be rephrased in terms of geometric constraints and study its implications. It is interesting to note an analogy with the $T=1$ models that can be realized as compactifications of the heterotic string on orbifolds of $T^4$. Also in that case, compactification degrees of freedom can contribute to the gauge symmetry due to stringy effects arising in the small volume region of moduli space. In that context, this can lead to $c^{gauge}_L>16$, which indeed occurs for some of the asymmetric orbifold models studied in \cite{BaykaraHamadaTaraziVafa2023,Baykara:2024vss}.

Looking at the list of anomaly-free $T=0$ candidate supergravity theories in \cite{HamadaLoges2023}, one finds that a large fraction  indeed has $c_L^{gauge} > 24$.  A natural question to pose at this point is what are the highest values of $c_L^{gauge}$ that can be attained, or in a sense how much can the worldsheet theory deviate from a geometric NLSM? It turns out that within the database of \cite{HamadaLoges2023} there exist five distinct theories for which $31 < c_L^{gauge} < 32$, including the $A_7$ model with matter in the $\mathbf{336}$ `box' representation \cite{Kumar:2010am,KimShiuVafa2019}. A feature which is shared by all these models is that they possess no neutral hypermultiplets in their spectrum, which suggests that they may have been been obtained by going to small volume limit of the $\mathbb{P}^2$ according to the picture suggested in \cite{BaykaraHamadaTaraziVafa2023}. Moreover, all these theories violate the Kodaira condition.

These models are reported in Table \ref{tab:clg>=31models}, along with the value of the residual central charge $c_L^{res} = 32 - c_L^{gauge}$. As $c_L^{res} < 1$, unitarity of the string should require the central charge to correspond to the central charge of a unitary  Virasoro  minimal model $\mathcal{M}(p+1,p)$:
\begin{equation}
c^{\mathcal{M}(p+1,p)} = 1- \frac{6}{p(p+1)}.
\end{equation}
We view this as a new type of non-trivial consistency check based on unitarity of the $L$-string worldsheet that allows us to test whether this class of theories belongs to the swampland or not. Interestingly, this simple check turns out to be satisfied in each and every one of entries of Table \ref{tab:clg>=31models}, for which the residual central charges turn out to be respectively those of the $\mathcal{M}(4,3),$ $\mathcal{M}(4,3),$ $\mathcal{M}(5,4),$ $\mathcal{M}(6,5),$ and $\mathcal{M}(7,6)$ models.
\begin{table}[H]
\centering
\begin{tabular}{c|c|c|c}
\textbf{Current algebra} & \textbf{Charged Matter} $\mathcal{H}$ & \textbf{$c^{gauge}_L$} & \textbf{$c^{res}_L$} \\
\hline
$A_{7, 8}$  & $\mathbf{336}$ & $63/2$ & 1/2 \\
$A_{5,6} \oplus D_{4, 6}$  & $(\mathbf{56},\mathbf{1}) +  (\frac{1}{2}\mathbf{20}, \mathbf{28})$ &  $63/2$ & 1/2\\
$A_{3,6} \oplus A_{3,6} \oplus E_{7,2}$  & $(\mathbf{10},\mathbf{10}, \mathbf{1}) + (\mathbf{1}, \mathbf{6},\frac{1}{2}\mathbf{56}) + (\mathbf{6}, \mathbf{1},\frac{1}{2}\mathbf{56})$ & $313/10$ & $7/10$\\
$E_{6, 8}$ & $\mathbf{351}'$ & $156/5$ &  4/5 \\
$A_{3,10} \oplus A_{11,2}$ & $(\mathbf{35},\mathbf{1}) +  (\mathbf{6}, \mathbf{66})$ & $218/7$ & 6/7
\end{tabular}
\caption{List of 6d anomaly-free theories with $c_L^{gauge} \ge 31$ where the factors $\frac{1}{2}$ stand for half-hypermultiplet representations.}
\label{tab:clg>=31models}
\end{table}

A striking prediction for these theories is that their $L$-string is evidently described by a rational conformal field theory with a finite spectrum of primaries! This is quite reminiscent of the description of K3 sigma models at Gepner points in the moduli space  in terms of $\mathcal{N}=2$ minimal models \cite{Gepner:1987qi,Eguchi:1988vra,Nahm:1999ps}, and raises the tantalizing possibility that one may be able to obtain a complete solution for these theories -- not just for their left moving sector. This possibility is currently under investigation \cite{wip} and we hope to report on it in future works.\\

In this paper, we will exploit rationality of the CFT to compute the elliptic genus for one of these models, namely the $\mathcal{T}_{E_6,\mathbf{351'},0}$. In Section \ref{sec:E6string} we will be able to determine its decomposition in terms of $E_{6,8}$ and three-state Potts model characters, and will find that in the unflavored limit the elliptic genus reduces to the one of the maximally Higgsed model $\mathcal{T}_{\emptyset,\mathbf{0},273}$ considered in \cite{HuangKatzKlemm2015,HaghighatMurthyVafaVandoren2015}. We view this as evidence that this 6d theory, which is non-geometric, does not belong to the swampland and that it can be progressively Higgsed all the way down to the theory with no gauge symmetry, following a path that crosses over from a non-geometric to a geometric region.\footnote{ In principle, the fact that the unflavored elliptic genus is the same at the top and at the bottom of the Higgsing transitions by itself does not imply that it is constant all the way along the path. A counterexample is the $E_8\times E_8$ heterotic string on K3, where one can move between configurations corresponding to different instanton numbers on the two $E_8$ factors via small instanton transitions. During these transitions the elliptic genus changes as the heterotic string becomes decomposable into two $E$-strings, and at then returns to the original value at the end of the process. However, we do not believe that this scenario is realized in our case as the Higgsing flow does not change the rank of the string charge lattice and we expect the BPS string to remain primitive throughout the path.}

In light of the discussion of Section \ref{subsec:Higgs}, we propose that it is possible to progressively Higgs the gauge group going through the various entries of Table \ref{tab:higgsingchainmodels}, which is repeated here as Table \ref{tab:higgsing2} for convenience, with information about the central charge $c^{gauge}_L$ added. The transition between non-geometric and geometric models occurs after Higgsing from $G_2$ to $A_2$, as discussed in Section \ref{subsec:Higgs}.  Here we see explicitly that $c_L^{gauge}<24$ is certainly not a sufficient condition for a model to admit a geometric realization, as can be seen by considering for example the non-geometric $G_2$ model for which $c^{gauge}_L \simeq 9.33\ll 24$.

Note that the affine algebras before and after a Higgsing transition are not in general related by a conformal embedding. It would be interesting to understand the individual Higgsing transitions better from the point of view of the worldsheet theory of the string; we leave this problem to future work.\\

In the next section we present some general considerations on the supersymmetric spectrum of hyperplane strings and its relation to the 6d massless spectrum, after which in Section \ref{sec:E6string} we turn to a detailed study of the $L$-string at the $E_{6,8}$ locus in $\mathcal{M}_{2d}^{[L]}$ and of its elliptic genus.

\begin{table}[H]
\centering
\begin{tabular}{c|c|c|c}
{$G$} & $H_0$ &$\mathcal{H}$ & $c^{gauge}_{L}$ \\
\hline\hline
$E_6$ & $0$ & $\mathbf{351}'$ & $156/5=31.2$ \\
%\hline
$F_4$ & $1$ & $\mathbf{324}$ & $416/17 \approx 24.47$ \\
%\hline
$B_4$ & $2$ & $\mathbf{9} + \mathbf{44} + \mathbf{126} + \mathbf{128}$ & $96/5= 19.2$ \\
%\hline
$D_4$ & $4$ & $\mathbf{8}_v + \mathbf{8}_s + \mathbf{8}_c + \mathbf{35}_v + \mathbf{35}_s + \mathbf{35}_c + \mathbf{56}_v + \mathbf{56}_s + \mathbf{56}_c$ & $16$ \\
%\hline
$B_3$ & $6$ & $ \mathbf{7} + 4 \times \mathbf{8} + \mathbf{21} +  \mathbf{27} + 3 \times  \mathbf{35} + 2 \times  \mathbf{48}$ & $168/13\approx 12.92$ \\
%\hline
$G_2$ & $13$ & $10 \times \mathbf{7} + 3 \times \mathbf{14} + 6 \times  \mathbf{27}$ & $28/3\approx 9.33$ \\
%\hline 
$A_2$ & $29$ &  $36 \times \mathbf{3} + 9 \times \mathbf{8} + 12 \times \mathbf{6}$
& $64/11 \approx 5.82$ \\
%\hline
$A_1$ & $85$ & $21 \times  \mathbf{3} + 64 \times  \mathbf{2}$ & $12/5  = 2.4$ \\
%\hline
$\emptyset$ & $273$ & No charged matter & $0$ \\
%\hline
\end{tabular}
\caption{Spectra of some 6d anomaly-free theories and the contribution \textbf{$c_L^{gauge}$} of the affine Lie algebra to $\textbf{$c_L$}$ of the hyperplane string CFT.}
\label{tab:higgsing2}
\end{table}
\subsection{The elliptic genus and the 6d massless spectrum}
\label{sec:ell}
The elliptic genus of a 2d $(0,4)$ CFT is defined as
\begin{equation}
\mathbb{E}(\vec{m},\tau) = \text{Tr}_R(-1)^F q^{H_L}\overline{q}^{H_R} e^{\vec{m}\cdot\vec{J}},
\end{equation}
where $F$ is the right-moving fermion number, $H_{L,R}$ are respectively the left- and right-moving Hamiltonians, $q=e^{2\pi i \tau}$, $\vec{J}$ are currents in the Cartan of the flavor symmetry group, and $\vec{m}$ are their conjugate chemical potentials. In our main example, $\vec{J}$ are the generators of the Cartan of $E_6$ and we will denote the chemical potentials by $\vec{m}_{E_6}$. 
For a Higgsing transition to take place, chemical potentials corresponding to the Cartan elements that get Higgsed must be tuned to zero as the currents decouple (this corresponds to turning off Kähler parameters associated to exceptional fibral divisors, after which complex structure deformations of the elliptic threefold can be turned on to reduce the singularity of the Kodaira fiber).

In the computations that follow, it will be helpful to also consider the \emph{unflavored} elliptic genus
\begin{equation}
\mathbb{E}(\tau) = \mathbb{E}(0,\tau) = \text{Tr}_R(-1)^F q^{H_L}\overline{q}^{H_R},
\end{equation}
which we expect to be an invariant under Higgsing. When the 2d CFT admits a description as a nonlinear sigma model, the elliptic genus is also invariant under Kähler deformations of the target space. In particular, this leads us to expect that if the $\mathcal{T}_{E_{6},\mathbf{351'},0}$ model is path-connected to $\mathcal{T}_{\emptyset,\mathbf{0},273}$, we should have
\begin{equation}
\mathbb{E}(\tau) = \frac{E_4(\tau)(31E_4(\tau)^3+113 E_6(\tau)^2)}{48\eta(\tau)^{32}},
\end{equation}
which is the value of the elliptic genus in the fully Higgsed model \cite{HuangKatzKlemm2015}.
The elliptic genus captures the spectrum of supersymmetric excitations of the string, which are ground states of the right-moving Hamiltonian and obey a quantization condition on the left-moving side:
\begin{equation}
H_L\vert\psi\rangle = p\vert\psi\rangle,\qquad p\in \mathbb{Z}-\frac{c_L}{24}
\end{equation}
From the target space point of view, the states that contribute have the interpretation as excitations of a string which wraps once around a torus and carries $p$ units of momentum \cite{Dijkgraaf:1996xw}. 

As a consequence, the elliptic genus admits a $q$-series expansion with integer grading:
\begin{equation}
\mathbb{E}(\vec{m},\tau) = \sum_{k\geq 0}C_k(\vec{m}) q^{-\frac{c_L}{24}+k},
\end{equation}
where $ C_k(\vec{m}) $ is the character of the (virtual) representation of the chiral symmetry group under which the $p=k$ states transform.

In the more general context of BPS strings of 6d theories, the spectrum of excitations contributing to the elliptic genus can sometime have a clear interpretation in terms of the massless spectrum of the parent theory. This has been discussed before in the case of the heterotic string \cite{Lee:2018urn, Hayashi:2023hqa}. In this case, the massless particle spectrum can be seen in the elliptic genus by looking at excitations at energy level $H_L = 0$. The corresponding coefficient of the elliptic genus is given by the chiral index
\begin{equation}
2(V-H_{\text{tot}}+T+3).
\end{equation}

Interestingly, the massless spectrum also makes an appearance in the elliptic genus of the hyperplane string, as can be checked in a number of examples \cite{HuangKatzKlemm2015,sjunp,Hayashi:2023hqa}. Namely, one observes that:
\begin{equation}
C_1 = 2 \mathcal{I}(G,\mathcal{H}, H_0) - 4 C_0
\label{eq:c1com}
\end{equation}
where we have subtracted a contribution $4 C_0$ coming from the center of mass factor
\begin{equation}
\mathbb{E}_{c.m.}(\tau) = \frac{1}{\eta(\tau)^4} = 1+4\,q+\mathcal{O}(q^2)
\end{equation}
which we have stripped off from the elliptic genus.

A heuristic way to understand this phenomenon which does not depend on geometric considerations is that a string wrapped on a torus can form bound states at threshold with massless BPS particles, in which case they are expected to appear as excited modes of the string \cite{Tong:2008qd}. For geometric theories the degeneracy by which these states contribute to the elliptic genus turns out to be 2. This can be understood as follows: consider the case where the 6d gauge symmetry is trivial; upon compactification from 6d to 5d the relevant states are interpreted as M2 branes wrapping a curve
\begin{equation}
[C] = [L]+[E]
\end{equation}
in the elliptic threefold, where $E$ denotes the elliptic fiber. The multiplicity of the states that contribute is given by the BPS invariant for this curve class, which corresponds up to a sign to the Euler characteristic of the moduli space of rank one sheaves on $[C]$. The essential elements of this computation have been discussed in \cite{HuangKatzKlemm2015}, and the upshot is that the answer includes a factor of minus the Euler characteristic of the threefold, which coincides with the chiral index of massless particles, as well as a factor of 
\begin{equation}
\chi(\mathbb{P}^1) = 2,
\end{equation}
where the $\mathbb{P}^1$ parametrizes the point on the hyperplane curve where the elliptic fiber attaches to it. Even when the elliptic fiber is allowed to degenerate, this component of the moduli space of sheaves is expected to be present, so that the geometric argument accounts for the factor of two in that case as well (at least for trivial Mordell-Weyl group). It would certainly be very interesting to have a deeper understanding of this phenomenon independent of geometric considerations, but we will not address the question further here.

\section{Exact solution for the hyperplane string at the $G=E_6$ locus}
\label{sec:E6string}

We now switch gears and perform a non-trivial test of our conjectures by going to the locus on the moduli space of the string $\mathcal{M}^{[L]}_{2d}$ corresponding to the 6d theory $\mathcal{T}_{E_{6},\mathbf{351}',0}$. We find that the left moving sector of its $L$-string is given by the following rational conformal field theory:
\begin{equation}
E_{6,8}\times \text{three state Potts model}.
\label{eq:rcftdec}
\end{equation}
and determine its elliptic genus.\\

\noindent We arrive at these results by combining the following properties:
\begin{enumerate}
\item Decomposition in terms of extended RCFT characters:
\begin{equation}
\mathbb{E}(\vec{m}_{E_6},\tau) = \sum_\alpha \chi^{Potts}_\alpha(\tau) \widetilde{\chi}^{E_{6,8}}_\alpha(\vec{m}_{E_6},\tau),
\label{eq:eg}
\end{equation}
where $\widetilde{\chi}^{E_{6,8}}(\vec{m}_{E_6},\tau)$ are linear combinations of affine $E_{6}$ characters at level 8
\begin{equation}
\widetilde{\chi}^{E_{6,8}}_\alpha(\vec{m}_{E_6},\tau) = \sum_{\lambda}n_{\alpha,\lambda} \chi^{E_{6,8}}_{\lambda}(\vec{m}_{E_6},\tau)
\label{eq:e6ans}
\end{equation}
with integer coefficients $n_{\alpha,\lambda}$ to be determined;
\item Modularity:
\begin{equation}
\mathbb{E}(-1/\tau) = \mathbb{E}(\tau);
\end{equation}
\item Relation to 6d massless spectrum \eqref{eq:c1com}:\footnote{ We have assumed that the relation to the 6d chiral index involves a factor of two for non-geometric theories as well as for the geometric ones. If we did not assume this, we would still be able to fully determine the elliptic genus up to an overall numerical factor.}
\begin{equation}
\mathbb{E}(\vec{m}_{E_6},\tau)\bigg\vert_{q^{-1/3}}
=
2\mathcal{I}(E_6,\mathbf{351'}, 0)-4 n_{vac}
=
4\cdot \mathbf{78}- 2\cdot (\mathbf{351'}+\mathbf{\overline{351}'})+12-4n_{vac},
\end{equation}
where $n_{vac}$ is the number of vacua of the theory, corresponding to the $q^{-4/3}$ coefficient in the elliptic genus.
\end{enumerate}
Remarkably it turns out that these three basic properties are enough  to uniquely determine the elliptic genus and its decomposition in terms of $E_{6,8}$ characters. We turn to this computation next, while in Section \ref{eq:ellcomm} we comment on some surprising features of the result.

\subsection{Computation of the elliptic genus}
\label{sec:computation}

Note that in order to use modularity we find it convenient to set the $ E_6$ parameters to zero and consider the unrefined elliptic genus $\mathbb{E}(\tau)=\mathbb{E}(\vec{m}_{E_6}=\vec{0},\tau)$. This turns out to be sufficient to achieve our goals. The constraints arising from properties 1.) and 2.) are discussed respectively in appendices \ref{app:consaff} and \ref{app:consmod} and are summarized here.\\

From property 1.) we obtain the following expansion for the flavored characters:
\begin{align}
\widetilde{\chi}^{E_{6,8}}_1
&\simeq
n_{1,\mathbf{1}}q^{-\frac{13}{10}}(\mathbf{1}+\mathbf{78}q)+\mathcal{O}(q^{\frac{7}{10}});
\\
\nonumber
\widetilde{\chi}^{E_{6,8}}_\epsilon 
&\simeq
n_{\epsilon,\mathbf{78}}\mathbf{78}q^{-\frac{7}{10}}
+
(\mathbf{34749}n_{\epsilon,\mathbf{34749}}+
(\mathbf{1}+
\mathbf{78}+
\mathbf{650}+
\mathbf{2430}+
\mathbf{2925}
)n_{\epsilon,\mathbf{78}})q^{\frac{3}{10}}\\
&+
\mathcal{O}(q^{\frac{13}{10}});\\
\widetilde{\chi}^{E_{6,8}}_{\psi_1}
&\simeq
n_{\psi_1,\mathbf{7371}}\mathbf{7371}q^{\frac{1}{30}}+\mathcal{O}(q^{\frac{31}{30}});\\
\widetilde{\chi}^{E_{6,8}}_{\psi_2}
&\simeq
n_{\psi_2,\mathbf{\overline{7371}}}\mathbf{\overline{7371}}q^{\frac{1}{30}}+\mathcal{O}(q^{\frac{31}{30}});
\\
\nonumber
\widetilde{\chi}^{E_{6,8}}_{\sigma_1}
&\simeq
n_{\sigma_1,\mathbf{\overline{351}'}}\mathbf{\overline{351}'}q^{-\frac{11}{30}}
+
(\mathbf{\overline{54054}} n_{\epsilon,\mathbf{\overline{54054}}}
+
(
\mathbf{\overline{351}}+
\mathbf{\overline{351}'}+
\mathbf{\overline{7371}}+
\mathbf{\overline{19305}}) n_{\epsilon,\mathbf{351}'})q^{\frac{19}{30}}
\\
&+
\mathcal{O}(q^{\frac{49}{30}});
\\
\nonumber
\widetilde{\chi}^{E_{6,8}}_{\sigma_2}
&\simeq 
n_{\sigma_2,\mathbf{{351'}}}\mathbf{{351}'}q^{-\frac{11}{30}}
+
(\mathbf{54054} n_{\epsilon,\mathbf{54054}}+
(
\mathbf{351}+
\mathbf{351'}+
\mathbf{7371}+
\mathbf{19305}) n_{\epsilon,\mathbf{351}'})q^{\frac{19}{30}}
\\
&+
\mathcal{O}(q^{\frac{49}{30}}).
\end{align}
On the other hand, from property 2.), we find that the $E_6$ part of the elliptic genus can be expressed as a linear combination of vector-valued modular forms (given in Equations \eqref{eq:v1}-\eqref{eq:v4}):
\begin{align}
\widetilde{\chi}^{E_{6,8}}(0,\tau)
&
=
(\alpha_0 (j-744) + \alpha_1)v_1(\tau)
+
\alpha_2 v_2(\tau)
+
\alpha_3 v_3(\tau)
+
\alpha_4 v_4(\tau)
\label{eq:vvmf}
\end{align}
with the following $q$-expansion:
\begin{align}
\nonumber
&
\widetilde{\chi}^{E_{6,8}}(0,\tau)
=
\\
&\begin{pmatrix}
\alpha_0 q^{-\frac{13}{10}}+(60\alpha_0+\alpha_1)q^{-\frac{3}{10}}+\mathcal{O}(q^{\frac{7}{10}})\\
(26\alpha_0+\alpha_2) q^{-\frac{7}{10}}+
(507\alpha_0+26\alpha_1-126\alpha_2+702\alpha_3-27\alpha_4)
q^{\frac{3}{10}}+\mathcal{O}(q^{\frac{13}{10}})\\
(6\alpha_0+\alpha_3) q^{-\frac{29}{30}}+\mathcal{O}(q^{\frac{1}{30}})\\
(156\alpha_0+\alpha_4) q^{-\frac{11}{30}}+
(2262\alpha_0+156\alpha_1-2310\alpha_2+8645\alpha_3+92\alpha_4)
q^{\frac{19}{30}}+\mathcal{O}(q^{\frac{49}{30}})\\
\end{pmatrix}
.
\label{eq:modanstext}
\end{align}

Moreover, from property 2.) we also learn that as a representation of $SL(2,\mathbb{Z})$ the characters \eqref{eq:modanstext} are invariant under charge conjugation. This means that the spectrum contributing to the elliptic genus is explicitly invariant under the $\mathbb{Z}_2$ symmetry that exchanges conjugate $E_6$ representations, so that
\begin{equation}
n_{\alpha,\mathbf{R}} = n_{\alpha,\mathbf{\overline{R}}}
\end{equation}
and we can write
\begin{align}
&
\widetilde{\chi}^{E_{6,8}}(0,\tau)
=
\begin{pmatrix}
\nonumber
\widetilde{\chi}^{E_{6,8}}_1(0,\tau)\\
\widetilde{\chi}^{E_{6,8}}_\epsilon(0,\tau)\\
\widetilde{\chi}^{E_{6,8}}_{\psi_1}(0,\tau) + \widetilde{\chi}^{E_{6,8}}_{\psi_2}(0,\tau)\\
\widetilde{\chi}^{E_{6,8}}_{\sigma_1}(0,\tau) + \widetilde{\chi}^{E_{6,8}}_{\sigma_2}(0,\tau)\\
\end{pmatrix}
\\
&
=
\begin{pmatrix}
n_{1,\mathbf{1}}q^{-\frac{13}{10}} + 78 n_{1,\mathbf{1}} q^{-\frac{3}{10}}+\mathcal{O}(q^{\frac{7}{10}})\\
78 n_{\epsilon,\mathbf{78}}q^{-\frac{7}{10}}+ (34749n_{\epsilon,\mathbf{34749}}+6084 n_{\epsilon,\mathbf{78}})q^{\frac{3}{10}}+\mathcal{O}(q^{\frac{13}{10}})\\
14742 n_{\psi_1,\mathbf{7371}}
q^{\frac{1}{30}}+\mathcal{O}(q^{\frac{31}{30}})\\
702 n_{\sigma_1,\mathbf{\overline{351}'}}q^{-\frac{11}{30}}
+ (108108 n_{\epsilon,\mathbf{54504}}+54756 n_{\epsilon,\mathbf{351}'})q^{\frac{19}{30}}
+
\mathcal{O}(q^{\frac{49}{30}})
\end{pmatrix}
.
\label{eq:affanstext}
\end{align}
Comparing Equation \eqref{eq:modanstext} and \eqref{eq:affanstext} we immediately find:
\begin{align}
\alpha_0&=n_{1,\mathbf{1}},\qquad
\alpha_1=18n_{1,\mathbf{1}},\qquad
\alpha_3=-6n_{1,\mathbf{1}},\\
\alpha_2&=26n_{1,\mathbf{1}}-78n_{\epsilon,\mathbf{78}},\qquad \alpha_4=-156n_{1,\mathbf{1}}+702n_{\sigma_1,\mathbf{351'}},
\end{align}
while Property 3.), using the explicit $q$-expansion of the Potts model characters obtained from \eqref{eq:chipotts}, determines the remanining coefficients:
\begin{equation}
n_{1,\mathbf{1}} = 3\qquad
n_{\epsilon,\mathbf{78}}= 1,
\qquad
n_{\sigma_1,\mathbf{351'}} = -2.
\end{equation}
\noindent This completely fixes the expression for the unflavored characters in terms of vvmf's as:
\begin{align}
\nonumber
\widetilde{\chi}^{E_{6,8}}(0,\tau)
&
=
(3 (j-744) + 54)v_1(\tau)
-
18 v_3(\tau)
-
1872 v_4(\tau)\\
&
=
\begin{pmatrix}
3 q^{-\frac{13}{10}}+234 q^{-\frac{3}{10}}+94770 q^{\frac{7}{10}}+78406000 q^{\frac{27}{10}}+ \mathcal{O}(q^{\frac{27}{10}})\\
78 q^{-\frac{7}{10}}+40833 q^{\frac{3}{10}}+ 10318266 q^{\frac{13}{10}}+\mathcal{O}(q^{\frac{23}{10}})\\
14742 q^{\frac{1}{30}}+5690412 q^{\frac{31}{30}}+695380140 q^{\frac{61}{30}}+\mathcal{O}(q^{\frac{91}{30}})\\
-1404 q^{-\frac{11}{30}}-1404q^{\frac{19}{30}}+118998298 q^{\frac{49}{30}}+\mathcal{O}(q^{\frac{19}{30}})\\
\end{pmatrix}
\label{eq:vvmf}
\end{align}
and we find that to high orders in the $q$-expansion
\begin{equation}
\mathbb{E}(0,\tau) = \sum_\alpha \chi^{Potts}_\alpha(\tau) \widetilde{\chi}^{E_{6,8}}_\alpha(0,\tau) = \frac{E_4(\tau)(31E_4(\tau)^3+113E_6(\tau)^2)}{48\eta(\tau)^{32}},
\label{eq:eg}
\end{equation}
in perfect agreement with the elliptic genus for the fully Higgsed theory!\footnote{We have checked this statement by expanding up to $\mathcal{O}(q^{10})$.} This supports our picture that the $\mathcal{T}_{E_6,\mathbf{351'},0}$ theory is realized at a locus on the moduli space of the CFT which is path-connected on ${\mathcal{M}}_{2d}^{[L]}$ to the trivial theory $\mathcal{T}_{\emptyset,\mathbf{0},273}$.

Proceeding in the same fashion and requiring that the remaining multiplicities $n_{\alpha,\mathbf{R}}$ in the expansion of the characters \eqref{eq:affanstext} all take small integer values leads to identify a unique plausible decomposition of the elliptic genus in terms of characters of the left moving RCFT \eqref{eq:rcftdec}, which we have checked continues to hold if we continue to expand the expressions to higher order. 

After the dust settles, we find the following expression for the $E_6$ part of the elliptic genus:
\begin{align}
\nonumber
\widetilde{\chi}^{E_{6,8}}_1(\vec{m}_{E_6},\tau)
=&
\phantom{+}
3\chi^{E_{6,8}}_{\mathbf{1}}
+\chi^{E_{6,8}}_{\mathbf{85293}}
-4\chi^{E_{6,8}}_{\mathbf{537966}}
+2\chi^{E_{6,8}}_{\mathbf{1559376}}
+2\chi^{E_{6,8}}_{\mathbf{\overline{1559376}}}
+3\chi^{E_{6,8}}_{\mathbf{3309696}} \\
&
\nonumber
+3\chi^{E_{6,8}}_{\mathbf{\overline{3309696}}}
+3\chi^{E_{6,8}}_{\mathbf{12514788}}
+\chi^{E_{6,8}}_{\mathbf{15359058}}
+\chi^{E_{6,8}}_{\mathbf{\overline{15359058}}}
+2\chi^{E_{6,8}}_{\mathbf{89791416}} \\
&
+2\chi^{E_{6,8}}_{\mathbf{\overline{89791416}}}
+\chi^{E_{6,8}}_{\mathbf{226459233}},
\label{eq:chiE61}
\\
\nonumber
\widetilde{\chi}^{E_{6,8}}_\epsilon(\vec{m}_{E_6},\tau)
=&
\phantom{+}
\chi^{E_{6,8}}_{\mathbf{78}}
+\chi^{E_{6,8}}_{\mathbf{34749}}
+3\chi^{E_{6,8}}_{\mathbf{2453814}}
-2\chi^{E_{6,8}}_{\mathbf{442442}}
-2\chi^{E_{6,8}}_{\mathbf{\overline{442442}}}
+4\chi^{E_{6,8}}_{\mathbf{64205141}}
\\
\nonumber
&
+2\chi^{E_{6,8}}_{\mathbf{47783736}}
+2\chi^{E_{6,8}}_{\mathbf{\overline{47783736}}}
+3\chi^{E_{6,8}}_{\mathbf{53557504}}
+3\chi^{E_{6,8}}_{\mathbf{\overline{53557504}}}
+3\chi^{E_{6,8}}_{\mathbf{29422393}}
\\
&
+\chi^{E_{6,8}}_{\mathbf{314269956}}
+\chi^{E_{6,8}}_{\mathbf{\overline{314269956}}},
\label{eq:chiE6epsilon}
\\
\nonumber
\widetilde{\chi}^{E_{6,8}}_{\psi_1}(\vec{m}_{E_6},\tau)
=
&
\phantom{+}
\chi^{E_{6,8}}_{\mathbf{7371}}
+2\chi^{E_{6,8}}_{\mathbf{393822}}
+3\chi^{E_{6,8}}_{\mathbf{494208}}
+\chi^{E_{6,8}}_{\mathbf{3675672}}
+3\chi^{E_{6,8}}_{\mathbf{5895396}}
+2\chi^{E_{6,8}}_{\mathbf{6675669}}
\\
\nonumber
&
+\chi^{E_{6,8}}_{\mathbf{17918901}}
+3\chi^{E_{6,8}}_{\mathbf{37459422}}
-4\chi^{E_{6,8}}_{\mathbf{41442192}}
+\chi^{E_{6,8}}_{\mathbf{64849356}}
+2\chi^{E_{6,8}}_{\mathbf{194548068'}}
\\
&
+2\chi^{E_{6,8}}_{\mathbf{310240854}}
+3\chi^{E_{6,8}}_{\mathbf{389321856}},
\label{eq:chiE6psi1}
\\
\nonumber
\widetilde{\chi}^{E_{6,8}}_{\psi_2}(\vec{m}_{E_6},\tau)
=
&
\phantom{+}
\chi^{E_{6,8}}_{\mathbf{\overline{7371}}}
+2\chi^{E_{6,8}}_{\mathbf{\overline{393822}}}
+3\chi^{E_{6,8}}_{\mathbf{\overline{494208}}}
+\chi^{E_{6,8}}_{\mathbf{\overline{3675672}}}
+3\chi^{E_{6,8}}_{\mathbf{\overline{5895396}}}
+2\chi^{E_{6,8}}_{\mathbf{\overline{6675669}}}
\\
\nonumber
&
+\chi^{E_{6,8}}_{\mathbf{\overline{17918901}}}
+3\chi^{E_{6,8}}_{\mathbf{\overline{37459422}}}
-4\chi^{E_{6,8}}_{\mathbf{\overline{41442192}}}
+\chi^{E_{6,8}}_{\mathbf{\overline{64849356}}}
+2\chi^{E_{6,8}}_{\mathbf{\overline{194548068}'}}
\\
&
+2\chi^{E_{6,8}}_{\mathbf{\overline{310240854}}}
+3\chi^{E_{6,8}}_{\mathbf{\overline{389321856}}},
\label{eq:chiE6psi2}
\\
\nonumber
\widetilde{\chi}^{E_{6,8}}_{\sigma_1}(\vec{m}_{E_6},\tau)
=&
\phantom{+}
-2\chi^{E_{6,8}}_{\mathbf{\overline{351}'}}
+\chi^{E_{6,8}}_{\mathbf{\overline{54054}}}
+\chi^{E_{6,8}}_{\mathbf{\overline{853281}}}
+3\chi^{E_{6,8}}_{\mathbf{\overline{4582656}}}
+2\chi^{E_{6,8}}_{\mathbf{\overline{6243237}}}
+4\chi^{E_{6,8}}_{\mathbf{\overline{7601958}}}
\\
\nonumber
&
+\chi^{E_{6,8}}_{\mathbf{\overline{30718116}}}
+3\chi^{E_{6,8}}_{\mathbf{\overline{93459366}}}
+3\chi^{E_{6,8}}_{\mathbf{\overline{159629184}}}
+2\chi^{E_{6,8}}_{\mathbf{\overline{194548068}}}
+3\chi^{E_{6,8}}_{\mathbf{\overline{219490128}}}
\\
&
+\chi^{E_{6,8}}_{\mathbf{\overline{32424678}}}
-2\chi^{E_{6,8}}_{\mathbf{\overline{70744752}}},
\label{eq:chiE6sigma1}
\\
\nonumber
\widetilde{\chi}^{E_{6,8}}_{\sigma_2}(\vec{m}_{E_6},\tau)
=
&
\phantom{+}
-2\chi^{E_{6,8}}_{\mathbf{351'}}
+\chi^{E_{6,8}}_{\mathbf{54054}}
+\chi^{E_{6,8}}_{\mathbf{853281}}
+3\chi^{E_{6,8}}_{\mathbf{4582656}}
+2\chi^{E_{6,8}}_{\mathbf{6243237}}
+4\chi^{E_{6,8}}_{\mathbf{7601958}}
\\
\nonumber
&
+\chi^{E_{6,8}}_{\mathbf{30718116}}
+3\chi^{E_{6,8}}_{\mathbf{93459366}}
+3\chi^{E_{6,8}}_{\mathbf{159629184}}
+2\chi^{E_{6,8}}_{\mathbf{194548068}}
+3\chi^{E_{6,8}}_{\mathbf{219490128}}
\\
&
+\chi^{E_{6,8}}_{\mathbf{32424678}}
-2\chi^{E_{6,8}}_{\mathbf{70744752}}.
\label{eq:chiE6sigma2}
\end{align}
\subsection{Remarks on symmetries of the elliptic genus}
\label{eq:ellcomm}

The result we have arrived at is remarkable: modularity of the elliptic genus (see in particular the discussion in Appendix \ref{app:consmod}) predicts that the combinations of characters \eqref{eq:chiE61}-\eqref{eq:chiE6sigma2} transform as a six-dimensional representation of $SL(2,\mathbb{Z})$ which decomposes into the four-dimensional irrep
\begin{align}
&
\widetilde{\chi}^{E_{6,8}}(\vec{m}_{E_6},\tau)
=
\begin{pmatrix}
\widetilde{\chi}^{E_{6,8}}_1(\vec{m}_{E_6},\tau)\\
\widetilde{\chi}^{E_{6,8}}_\epsilon(\vec{m}_{E_6},\tau)\\
\widetilde{\chi}^{E_{6,8}}_{\psi_1}(\vec{m}_{E_6},\tau)
+
\widetilde{\chi}^{E_{6,8}}_{\psi_2}(\vec{m}_{E_6},\tau)\\
\widetilde{\chi}^{E_{6,8}}_{\sigma_1}(\vec{m}_{E_6},\tau)
+
\widetilde{\chi}^{E_{6,8}}_{\sigma_2}(\vec{m}_{E_6},\tau)\\
\end{pmatrix}
,
\end{align}
which is invariant charge conjugation, and a two-dimensional irrep which transforms to minus itself under $\mathcal{C}$ given by
\begin{equation}
\widetilde{\chi}_{-}^{E_{6,8}}(\vec{m}_{E_6},\tau)
=
\begin{pmatrix}
\widetilde{\chi}^{E_{6,8}}_{\psi_1}
\!-\!
\widetilde{\chi}^{E_{6,8}}_{\psi_2}\\
\vspace{-.1in}\\
\widetilde{\chi}^{E_{6,8}}_{\sigma_1}
\!-\!
\widetilde{\chi}^{E_{6,8}}_{\sigma_2}
\end{pmatrix}
.
\label{eq:chiminus}
\end{equation}
Indeed, we have been able to explicitly verify that this is the case!\footnote{ We have established this fact in cooperation with Yann Proto and plan to provide further details and applications of this approach in a future publication \cite{wip}.}

The prediction of the existence of this modular covariant combination of characters is in itself highly nontrivial, and is reminiscent of constructions arising in the classification of modular invariant partition functions of WZW models \cite{Warner:1989yy,Roberts:1990tv,Gannon:1995jm,Gannon:1995uv}. It would be very interesting to understand if the methods developed in that context can give rise to this specific combination of characters and shed light on its significance.\\

Of the two irreps of $SL(2,\mathbb{Z})$, it is the four-dimensional representation that contributes to the elliptic genus, while the latter does not, even in the flavored case. The reason for this is simple: it couples to the two-dimensional representation of the Potts model with entries
\begin{equation}
{\chi}_{-}^{Potts}(\tau)
=
\begin{pmatrix}
\frac{1}{2}
(
{\chi}^{Potts}_{\psi_1}
\!-\!
{\chi}^{Potts}_{\psi_2}
)
\\
\vspace{-.1in}\\
\frac{1}{2}
(
{\chi}^{Potts}_{\sigma_1}
\!-\!
{\chi}^{Potts}_{\sigma_2}
)
\end{pmatrix}
,
\label{eq:chipottsminus}
\end{equation}
which is trivially zero since the character does not distinguish between the two charge conjugate representations $\psi_{1,2}$, and likewise for $\sigma_{1,2}$.

This raises the question if one can detect the contribution from \eqref{eq:chiminus} to the supersymmetric spectrum of the string by any other means. This can be achieved by computing the \emph{twining} genera (along the lines of \cite{Gaberdiel:2010ch,Eguchi:2010fg,Harrison:2013bya,Cheng:2015rby}):
\begin{equation}
\mathbb{E}_g(\vec{m}_{E_6},\tau) = \text{Tr}_R\, g\,(-1)^{F_R} q^{H_L}\overline{q}^{H_R} e^{\vec{J}_{E_6}\cdot \vec{m}_{E_6}}, \qquad 
\end{equation}
with the insertion of a generator $g$ of the $\mathbb{Z}_3$ symmetry of the Potts model. It would be interesting to see how the twining genus we propose relates, if at all, to the torsion-refined GV invariants of elliptic fibrations over $\mathbb{P}^2$ which have been introduced in \cite{Schimannek:2021pau}.

We do, in fact, have some indications suggesting that this contribution should appear: quite surprisingly, the multiplicities $n_{\alpha,\lambda}$ appearing in Equations \eqref{eq:chiE61}-\eqref{eq:chiE6sigma2} are such that the elliptic genus is invariant under a distinct $\mathbb{Z}_3$ outer automorphism symmetry rotating the external legs of the affine $E_6$ Dynkin diagram, in the sense that
\begin{equation}
\lambda' = \tilde{g} \circ \lambda \implies n_{\alpha,\lambda}=n_{\tilde{g}\circ\alpha,\lambda'}, \qquad \tilde{g}\in\mathbb{Z}_3,
\label{eq:rot}
\end{equation}
where $\lambda,\lambda'$ are affine $E_6$ Dynkin labels (see Appendix \ref{app:affine} for our conventions) and this $\mathbb{Z}_3$ acts on the Potts labels $\alpha$ as cyclic permutations of $(1,\psi_1,\psi_2)$ and $(\epsilon,\sigma_1,\sigma_2)$.\footnote{We remark that a similar symmetry appears in the (much less intricate!) triality-invariant model with $E_{6,1}$ symmetry discovered in \cite{Gadde:2014ppa}.} For example, this $\mathbb{Z}_3$ symmetry relates to each other the vacuum representation of $E_6$ at level 8, $\mathbf{1}$, with the two conjugate representations $\mathbf{5895396}$ and $\mathbf{\overline{5895396}}$, which reside respectively in the characters $\widetilde\chi^{E_{6,8}}_1,$ $\widetilde\chi^{E_{6,8}}_{\psi_1},$ and $\widetilde\chi^{E_{6,8}}_{\psi_2}$. Indeed, all three representations occur with multiplicity 3 in the elliptic genus. Keeping track of this symmetry, we find that we can write down a very concise expression for the elliptic genus:
\begin{align}
\nonumber
\mathbb{E}(\vec{m}_{E_6},\tau)
&=
\mathcal{O}^{Potts}_{1}
\cdot
(3\mathcal{O}^{(1)}_3\!+\!\mathcal{O}^{(4)}_3\!-\!4\mathcal{O}^{(5)}_3\!+\!3\mathcal{O}^{(8)}_3\!+\!\mathcal{O}^{(9)}_3
\!+\!\mathcal{O}^{(2)}_6\!+\!2\mathcal{O}^{(4)}_6\!+\!3\mathcal{O}^{(5)}_6\!+\!2\mathcal{O}^{(6)}_6)
\\
&
+
\mathcal{O}^{Potts}_{\epsilon}
\cdot
(\mathcal{O}^{(2)}_3\!+\!\mathcal{O}^{(3)}_3\!+\!3\mathcal{O}^{(6)}_3\!+\!4\mathcal{O}^{(7)}_3\!+\!3\mathcal{O}^{(10)}_3\!-\!
2\mathcal{O}^{(1)}_6\!+\!\mathcal{O}^{(3)}_6\!+\!3\mathcal{O}^{(7)}_6\!+\!2\mathcal{O}^{(8)}_6),
\end{align}
written in terms of the orbits of $E_6$ level 8 integrable highest weight representations under $S_3$, which are collected in Table \ref{tab:orbits} of Appendix \ref{app:consaff}. These are expressed as vectors $(\mathbf{R}_0,\mathbf{R}_1,\mathbf{R}_2)$ where the $\mathbf{R}_j$ component has charge $j$ under the $\mathbb{Z}_3$ center symmetry of $E_6$.
Likewise we have defined the three-component vectors
\begin{equation}
\mathcal{O}^{Potts}_{1}
=
\begin{pmatrix}
\chi_1^{Potts}\\
\chi_{\psi_1}^{Potts}\\
\chi_{\psi_2}^{Potts}
\end{pmatrix}
,
\qquad
\mathcal{O}^{Potts}_{\epsilon}
=
\begin{pmatrix}
\chi_\epsilon^{Potts}\\
\chi_{\sigma_2}^{Potts}\\
\chi_{\sigma_1}^{Potts}
\end{pmatrix}
\end{equation}
and a scalar product
\begin{equation}
\mathcal{O}^{Potts}_\alpha\cdot \mathcal{O}^{(i)}_n
=
\sum_{j=1}^3 (\mathcal{O}^{Potts}_\alpha)_j (\mathcal{O}^{(i)}_n)_j
\end{equation}
between them. Invariance under the $\mathbb{Z}_3$ outer automorphism symmetry \eqref{eq:rot} requires including the two-dimensional representations \eqref{eq:chiminus} and \eqref{eq:chipottsminus} in the elliptic genus. The appearance of this unexpected $\mathbb{Z}_3$ outer automorphism symmetry is currently under study \cite{wip}.

\section{Conclusions and future directions}

\label{sec:disc}

In this paper we have conjectured a relation between the quantum corrected moduli space of the hyperplane string CFT $\mathcal{M}_{2d}^{[L]}$ and the region ${\mathcal{M}}^{[L]}_{QG}$ of the landscape of  6d $\mathcal{N}=(1,0)$ quantum gravity theories where the string charge lattice has rank one. We have exploited this connection to explore regions of the landscape which cannot be described in terms of a geometric F-theory compactification.

In regions where the geometric realization is reliable, a $\mathcal{N}=(0,4)$ sigma model description for the worldsheet theory on the string is available, and its moduli space mirrors the  $\mathbb{P}^2$ base of the geometry and the K\"ahler volume of the base plays the role of a marginal deformation under which supersymmetry protected quantities like the elliptic genus are invariant. As one goes beyond the large volume region, quantum corrections are also expected to become important in the NLSM. Nonetheless, we posed the question of whether one can make sense of the quantum theory of the string and follow it into strong coupling regions as well. In particular we have asked about the fate of the 2d CFT in regions which from the perspective of the $L$-string's worldsheet look as non-geometric as possible. By this we mean that the contribution to the left-moving central charge coming from the current algebra that arises from the 6d gauge symmetry almost completely saturates $c_L$, leaving very little room for `geometric' degrees of freedom:
\begin{equation}
24 \ll c_L^{gauge} \leq 32.
\end{equation}
Surprisingly, we found five instances where the residual central charge $c^{res}_L = 32 - c_L^{gauge}$ is less than 1 and coincides with the central charge of a unitary Virasoro minimal model. The fact that the value $c_L^{res}$ is compatible with unitarity is in itself nontrivial evidence that the corresponding 6d theory might lie in the landscape. That these five theories are necessarily non-geometric is corroborated by the fact that their massless spectrum does not include any neutral hyper that could play the role of the universal hypermultiplet in a geometric compactification.

This simplification means that the powerful tools of rational conformal field theories are at our disposal to study these models. In this paper we have focused on a specific theory corresponding to $E_{6}$ gauge symmetry realized on the string as affine $E_6$ at level 8 and were able to determine its elliptic genus, which turns out to match the elliptic genus of geometric models in the unflavored limit. This supports the hypothesis that the $E_6$ model is connected to geometric models by a path in the region ${\mathcal{M}}^{[L]}_{QG}$ of the landscape, and in fact we have found a plausible Higgsing chain of 6d theories which are situated in the intermediate region.\\

\noindent There are several interesting directions which we deem worthy of future study:
\begin{itemize}
\item It would be very interesting to track the evolution of the elliptic genus along this sequence of Higgsing transitions.
\item Having determined the elliptic genus of the hyperplane string of the $E_{6,8}$ theory, it becomes of course an urgent task to test whether its spectrum is compatible with the (non-Abelian) Weak Gravity Conjecture along the lines of \cite{Lee:2018urn,Hayashi:2023hqa}.
\item The modular constraints imposed by RCFT are extremely constraining and raise the possibility of computing not only the elliptic genus, but also partition functions, the $(R,-)$ partition function being the obvious candidate as a starting point. While there is no reason to expect this to be feasible at generic points of moduli space, it may become possible at the special points we have identified, in a similar way that the K3 partition function becomes computable at Gepner points \cite{Eguchi:1988vra}. It would then become very interesting to interpret 2d observables from the perspective of the parent 6d theory. It would of course also be nice to gather additional information about other theories with $c_L^{gauge} > 31$. We have taken initial steps to analyze the $A_{7,8}\times \text{Ising}$ model (see Table \ref{tab:clg>=31models}) following the approach outlined in Section \ref{sec:computation} of this paper; while our Ansätze based on RCFT appear to be correct, we have not been able to fix the elliptic genus in this model unambiguously. On the other hand, a more rigorous approach based on explicitly constructing the relevant $SL(2,\mathbb{Z})$ irreducible representations should provide more powerful tools. We are currently investigating these issues in collaboration with Yann Proto and hope to report on them in future publications \cite{wip}.
\item From the VOA perspective, it is perhaps surprising that our explorations have led us to discover a four- and a two-dimensional irreducible representation built out of the 372 inequivalent characters of $E_{6,8}$. It would be very interesting to understand from a more fundamental point of view why these representations are singled out, and whether it can be constructed using some of the methods that were developed in the classification program of WZW modular invariants \cite{Warner:1989yy,Roberts:1990tv,Gannon:1995jm,Gannon:1995uv}.
\item While we have restricted our attention to regions of the landscape where the string charge lattice $\Gamma$ is one-dimensional, it would be very interesting to see whether it is possible to say something about regions where this does not hold. For instance, by a small-instanton transition one can arrive at models with one tensor multiplet which in F-theory are described by an elliptic fibration over $\mathbb{F}_1$. The small-instanton transition maps the hyperplane string to a BPS string which is no longer a primitive object; an open question is whether the string still retains a useful role as a tool to explore this region of the landscape. It would also be interesting to study situations in which some components of the discriminant locus have anomaly coefficient $b_i = 1$, for which at low energies the hyperplane string is expected to possess an instantonic branch of vacua in addition to the geometric branch.
\item While we have restricted our attention to regions of the landscape where the string charge lattice $\Gamma$ is one-dimensional, it would be very interesting to see whether it is possible to say something about regions where this does not hold. For instance, by a small instanton transition one can arrive at models with one tensor multiplet which in F-theory are described by an elliptic fibration over $\mathbb{F}_1$. The transition maps the hyperplane string to a BPS string which is no longer a primitive object. An open question is whether this string still retains a useful role as a tool to explore the landscape of theories with $T>1$. It would also be interesting to study situations in which some components of the discriminant locus have anomaly coefficient $b_i = 1$, for which at low energies the hyperplane string is expected to possess an instantonic branch of vacua in addition to the geometric branch.
\item Finally, we have noticed that all models with $c^{gauge}_L>31$ are at the same time rational and have $H_0 = 0$ neutral hypermultiplets in their spectrum.  It is natural to ask whether rationality is in fact a necessary feature for all theories with $H_0 =0$, including those for which $c_L^{gauge} \leq  31$, and more in general what is the criterion that determines whether a RCFT description is available. This would shed light on what is the range of applicability of the techniques we have begun to develop in the current paper.
\end{itemize}

\section*{Acknowledgements}
We are grateful to Michele Del Zotto, Hee-Cheol Kim, Ruben Minasian, Wati Taylor, and Cumrun Vafa for helpful discussions and especially to Yann Proto for collaboration on a related project. GL wishes to acknowledge the Erwin Schrödinger International Institute for Mathematics and Physics of the University of Vienna for hospitality during the  workshop ``The Landscape vs. the Swampland'', where this work was initiated. GL and LN also would like to acknowledge the ``Lotus \& Swamplandia 2025'' conference where it was completed. The work of GL and LN has received funding from the European Research Council (ERC) under the Horizon Europe (grant agreement No. 101078365) research and innovation program.

\appendix
\section{Modular forms: basic definitions}
\label{app:mf}
Throughout the paper we denote by $\tau$ the modular parameter and use $q = e(\tau)$ where $e(z):=e^{2\pi i z}$.

The Dedekind eta function is given by
\begin{equation}
\eta(\tau) = q^{\frac{1}{24}}\prod_{n=1}^\infty(1-q^n);
\end{equation}
and transforms with weight $1/2$ under $\tau\to-1/\tau$ as
 $ \eta(-1/\tau)=\sqrt{-i\tau}\eta(\tau)$. The Eisenstein series $E_{2k}(\tau) $ are defined in terms of the Bernoulli numbers $B_{k}$ and divisor sigma function $\sigma_k(n)$ as
\begin{equation}
E_{2k}(\tau) = 1- \frac{4k}{B_{2k}}\sum_{n=1}^\infty \sigma_{2k-1}(n)q^n
\end{equation}
and transform as modular forms of weight $2n$. The $j$-invariant is given by:
\begin{equation}
j(\tau) = \frac{E_4(\tau)^3}{\eta(\tau)^{24}} = q^{-1} +744+ 196884 q+21493760 q^2+864299970 q^3+\mathcal{O}(q^4).
\end{equation}

\section{RCFT data}
\label{app:rcft}
\subsection{The three-state Potts model}
\label{app:potts}
The $\mathcal{M}(6,5)$ Virasoro minimal model, which has central charge $c = 4/5$, possesses a finite spectrum of primaries $\phi_{r,s}$ labeled by two positive integers $r,s$ such that $r\leq 5, s\leq 4$ . As operators $\phi_{r,s}$ and $\phi_{6-r,5-s}$ are identified, it suffices to consider operators with $r\leq 5$ and $s \leq 2$. The conformal dimensions of this set of operators is given in Table \ref{tab:Potts}. The character associated to primary $\phi_{r,s}$ is given by:
\begin{equation}
\chi^{M(6,5)}_{r,s}(\tau) = \frac{1}{\eta(\tau)}\left(\sum_{n\in\mathbb{Z}}q^{\frac{(60n+5r-6s)^2}{120}}-\sum_{n\in\mathbb{Z}}q^{\frac{(60n+5r+6s)^2}{120}}\right).
\label{eq:chipotts}
\end{equation}

The three-state Potts model can be constructed starting from  $\mathcal{M}(6,5)$ by extending the Virasoro algebra, giving rise to a $W_3$ symmetry. Under the enlarged symmetry, the spectrum organizes into six conformal families $(1,\epsilon,\psi_{1,2},\sigma_{1,2})$ whose Virasoro primary content consists respectively of the following primaries:
\begin{align}
1:& \phi_{1,1}+\phi_{1,5}\\
\epsilon:& \phi_{2,1}+\phi_{2,5}\\
\psi_1:& \phi_{1,3}\\
\psi_2:& \phi_{1,3}\\
\sigma_1:& \phi_{2,3}\\
\sigma_2:& \phi_{2,3}
\end{align}
The six primaries have respectively conformal weights $0,2/5,2/3,2/3,1/15,1/15$. The Potts model characters, which we denote by $\chi^{Potts}_\alpha $ for primary $\alpha$, are given by combinations of characters of the Virasoro primaries. Crucially two copies each of the $\phi_{1,3}$ and $\phi_{2,3}$ appear in the spectrum which are related by charge conjugation and give rise to identical characters for $\psi_{1,2}$ as well as for $\sigma_{1,2}$.

The modular transformation matrices are given by:
\begin{align}
\mathcal{S}^{Potts}
&=
\frac{2}{\sqrt{15}}
\begin{pmatrix}
s_1& s_2 & s_1 & s_1 & s_2 & s_2\\
s_2& -s_1 & -s_2 & - s_2 & -s_1 & - s_1\\
s_1& s_2 & \omega s_1 & \overline\omega s_1 & \omega s_2& \overline\omega s_2\\
s_1& s_2 & \overline\omega s_1 & \omega s_1 & \overline\omega s_2& \omega s_2\\
s_2& -s_1 & \omega s_2& \overline\omega s_2& -\omega s_1 & -\overline\omega s_1\\
s_2& -s_1 & \overline\omega s_2& \omega s_2 & -\overline\omega s_1 & -\omega s_1
\end{pmatrix}\\
\mathcal{T}^{Potts}
&=
e(-\tfrac{1}{30})
\begin{pmatrix}
1& & & & &\\
& e(\tfrac{2}{5}) & & & &\\
& & e(\tfrac{2}{3}) & & &\\
& & & e(\tfrac{2}{3}) & &\\
& & & & e(\tfrac{1}{15})&\\
& & & & & e(\tfrac{1}{15})\\
\end{pmatrix},
\end{align}
where $s_\ell = \sin(\ell\pi/5)$ and $\omega = e(\tfrac{1}{3})$. These matrices satisfy the relations
\begin{equation}
(\mathcal{S}^{Potts})^2 = (\mathcal{S}^{Potts}\mathcal{T}^{Potts})^3 =\mathcal{C}^{Potts}
\end{equation}
where 

\begin{equation}
\mathcal{C}^{Potts}
=
\begin{pmatrix}

1& \phantom{111}&\phantom{111} &\phantom{111} &\phantom{111} &\phantom{111}\\
& 1 & & & &\\
& &  & 1 & &\\
& & 1 &  & &\\
& & & & & 1\\
& & & & 1& 
\end{pmatrix},
\qquad
(\mathcal{C}^{Potts})^2 = 1.
\end{equation}
The Potts model possesses a $\mathbb{Z}_3$ symmetry under which $1,\epsilon$ are neutral, $\psi_1 $ and $\sigma_1$ transform by a phase $e^{\frac{2\pi i}{3}}$ while $\psi_2$ and $\sigma_2$ transform with the opposite phase.

The two charge conjugate pairs of primaries have identical characters, which we denote by $\sigma_1$ and $\sigma_2$, and correspondingly the six-dimensional $SL(2,\mathbb{Z})$ representation presented above is reducible. The four non-trivial characters corresponding to the primaries $1,\epsilon, $ and to $ \psi =\frac{1}{2}(\psi_1+\psi_2) $ and $\frac{1}{2}(\sigma_1+\sigma_2)$ assemble into a vector valued modular form of
\begin{equation}
\chi^{Potts}(\tau)
=
\begin{pmatrix}
\chi^{Potts}_1\\
\chi^{Potts}_\epsilon\\
\chi^{Potts}_\psi\\
\chi^{Potts}_\sigma
\end{pmatrix}
(\tau)
,
\end{equation}
which transforms as a four-dimensional irreducible representation of $SL(2,\mathbb{Z})$ characterized by the following $S$- and $T$- matrices:
\begin{equation}
\widetilde{\mathcal{S}}^{Potts}
=
\frac{2}{\sqrt{15}}
\begin{pmatrix}
s_1& s_2 & 2s_1 & 2s_2\\
s_2& -s_1 & -2s_2 & -2s_1\\
s_1& s_2 & -s_1 & -s_2\\
s_2& -s_1 & -s_2 & s_1
\end{pmatrix},
\qquad
\widetilde{\mathcal{T}}^{Potts}
=
e(-\frac{1}{30})
\begin{pmatrix}
1& & &\\
& e(\tfrac{2}{5}) & &\\
& & e(\tfrac{2}{3}) &\\
& & & e(\tfrac{1}{15})
\end{pmatrix},
\end{equation}
for which $(\widetilde{\mathcal{S}}^{Potts})^2 = (\widetilde{\mathcal{S}}^{Potts}\widetilde{\mathcal{T}}^{Potts})^3 =1$.

The anti-symmetric combinations of operators $\sigma_1-\sigma_2$ and $\psi_1-\psi_2$, on the other hand, transform in a two-dimensional irreducible representation of the modular group and do not contribute to the elliptic genus because the corresponding combination of characters cancels out.

\begin{table}
\begin{center}
\begin{tabular}{c|ccccc}
s$\backslash$ r& 1&2&3&4&5\\
\hline
1 & $0$&$\frac{1}{8}$&$\frac{2}{3}$&$\frac{13}{8}$&$3$\\
2 & $\frac{2}{5}$&$\frac{1}{40}$&$\frac{1}{15}$&$\frac{21}{40}$&$\frac{7}{5}$
\end{tabular}
\end{center}
\caption{Conformal dimensions of Virasoro primaries $\phi_{r,s}$ within range $1 \leq r\leq 5$ and $1\leq s \leq 2$.}
\label{tab:Potts}
\end{table}

\subsection{Affine $E_{6}$ Lie algebra}
\label{app:affine}

\begin{figure}
    \centering
    \includegraphics[width=0.6\linewidth]{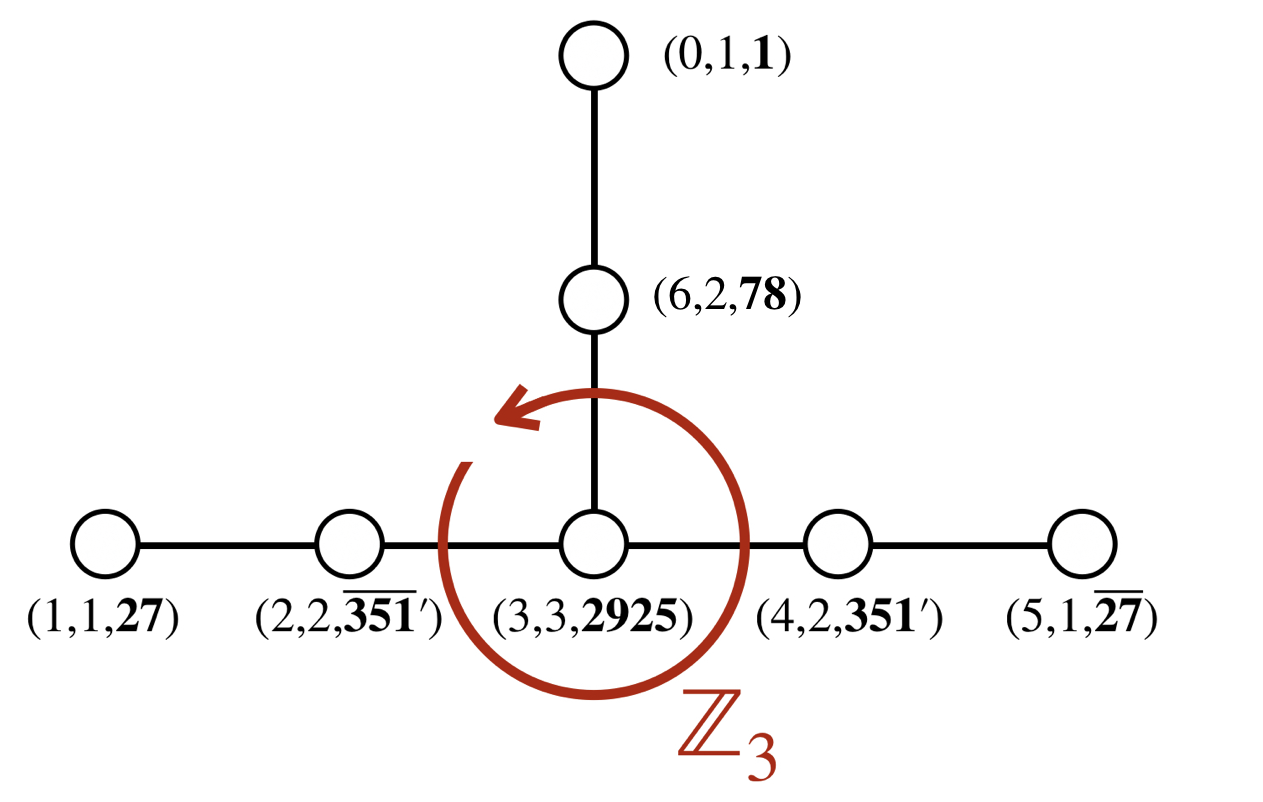}
    \caption{The $E_6$ affine Dynkin diagram. The quantities $(j,a^\vee,\mathbf{R})$ associated to each node denote respectively its label, its comark, and the associated $E_6$ representation. Depicted here are also the actions of the $\mathbb{Z}_2$ charge conjugation symmetry and $\mathbb{Z}_3$ outer automorphism group, which together assemble into the $D(E_6) = S_3$ diagram automorphism group of affine $E_6$.}
    \label{fig:e6Dynkin}
\end{figure}

In this appendix we summarize our notation and basic facts about the affine Lie algebra $E_6$ at level 8 which are relevant to the paper. The Lie algebra possesses a collection of basic representations of dimension $1,27,27,78,351,351,2925$ associated to the nodes of the affine Dynkin diagram as depicted in Figure \ref{fig:e6Dynkin}, where the comarks are also indicated. A highest weight representation can be characterized by its Dynkin label $[\lambda_1,\dots,\lambda_6]$ whose entries are indexed by the corresponding node. In this paper we find it convenient to alternatively denote a representation by its dimension, e.g. the adjoint representation is denoted as $\mathbf{78}$, and when there exist multiple representations we use the conventions of \cite{Feger:2019tvk}, e.g. the representations $[0,0,0,1,0,0],\, [0,1,0,0,0,0],\, [0,0,0,0,2,0] $, and $[2,0,0,0,0,0]$ are respectively denoted by $\mathbf{351},\mathbf{\overline{351}},\mathbf{351'},\mathbf{\overline{351}'}$. A representation that is obtained from representation $\mathbf{R}$ by the action of the $\mathbb{Z}_2$ outer automorphism group of $E_6$ is denoted by $\mathbf{\overline{R}}$. By a slight abuse of notation, given a representation $\mathbf{R}$ we also denote the corresponding flavored character, which depends on a choice of parameter $\vec{m}_{E_6}$ valued in the dual of the Cartan of $E_6$ by the same symbol.

For the affine Lie algebra associated to $E_6$ we denote representations in terms of their affine Dynkin labels $[\lambda_0,\dots,\lambda_6]$ associated to the nodes of the affine Dynkin diagram as in Figure \ref{fig:e6Dynkin}, or alternatively by the highest weight representation $\mathbf{R}= [\lambda_1,\dots,\lambda_6]$ of the corresponding finite-dimensional Lie algebra. Level $k$ integrable highest weight representations satisfy the constraint
\begin{equation}
 \sum_{a=0}^{6} a_i\lambda_i = k
\end{equation}
where $a_i = (1,1,2,3,2,1,2)$ are the comarks. The $\mathbb{Z}_3$ center of $E_6$ acts as a symmetry on the integrable highest weight representations, under which a representation $\lambda$ transforms with charge
\begin{equation}
e^{\frac{2\pi i C(\lambda)}{3}},\qquad C(\lambda) = \lambda_1-\lambda_5-\lambda_2+\lambda_4 \text{ mod }3.
\end{equation}

In the level $k$ example which is relevant to the paper there are 372 inequivalent integrable highest weight representations, consisting of 32 real and 170 conjugate pairs of complex representations whose components are also exchanged by charge conjugation. In this paper the $\mathbb{Z}_3$ \emph{affine} outer automorphism group which permutes the three legs of the affine Dynkin diagram will also play a role. This combines with charge conjugation to give the larger symmetry group $D(E_6) = S_3$ of diagram automorphisms of affine $E_6$. Integrable highest weight representations organize respectively into 32 three-dimensional orbits and 46 six-dimensional orbits under $S_3$. 

The conformal weight of the primary operator associated to an integrable h.w.r. $\lambda$ is given by $h_\lambda= \frac{C_2(\lambda)}{h^\vee+k}$, where $C_2(\lambda)$ is the second Casimir invariant normalized so that evaluated on the adjoint representation it equals the dual Coxeter number $h^\vee$ of $E_6$, 
\begin{equation}
h_\lambda \in \mathbb{Z}/60.
\label{eq:z60}
\end{equation}

\section{Constraints on the $E_{6,8}$ string elliptic genus}
\label{app:cons}
In this section we discuss the constraints on the form of the elliptic genus which follow from Properties 1.) and 2.) of Section \ref{sec:E6string}.

\subsection{Constraints from affine symmetry}
\label{app:consaff}
We recall that the $L$-string at the $\mathcal{T}_{E_6,\mathbf{351'},0}$ is described by a product of an affine $E_{6,8}$ current algebra and a residual CFT of $c=4/5$, which corresponds to two possible choices of RCFT: the $\mathcal{M}(6,5)$ Virasoro minimal model and the three-state Potts model. As a first step in constraining its the elliptic genus, we must assess which one of the two is the relevant model. This can be determined by looking at which sets of $\mathcal{M}(6,5)$ and $E_{6,8}$ characters combine to give operators with integer conformal dimension, as required in the elliptic genus. A glance at Table \ref{tab:Potts} and Equation \eqref{eq:z60} clarifies that this is not possible for the operators $\phi_{(r,s)}$ with $s$ even, which singles out the three-state Potts model as the correct $c=4/5$ VOA contributing to the left-moving side. 
\begin{table}
\begin{center}
\begin{tabular}{c|c|r}
$\alpha$& $h_{\mathbf{R}}$ & $\mathbf{R}$ \\
\hline\hline
1& 0&
$\mathbf{1}$
\\
& 2&
$\mathbf{85293}$
\\
 & 3&
$\mathbf{537966},
\mathbf{1559376},
\mathbf{\overline{1559376}},
\mathbf{3309696},
\mathbf{\overline{3309696}},
\mathbf{12514788}$
\\
 & 4&
 $
 \mathbf{15359058},
\mathbf{\overline{15359058}},
\mathbf{89791416},
\mathbf{\overline{89791416}},
\mathbf{226459233}
$
\\
\hline
$\epsilon$&3/5&
$\mathbf{78}$
\\
&8/5&
$\mathbf{34749}$
\\
&13/5&
$\mathbf{2453814}$
\\
&18/5&
$\mathbf{442442}$,
$\mathbf{\overline{442442}}$,
$\mathbf{64205141}$,
$\mathbf{47783736}$,
$\mathbf{\overline{47783736}}$,
$\mathbf{53557504}$,
\\
& &
$\mathbf{\overline{53557504}}$,
$\mathbf{29422393}$
\\
&23/5&
$\mathbf{314269956}$,
$\mathbf{\overline{314269956}}$
\\
\hline
$\psi_1$&4/3&
$\mathbf{7371}$
\\
&7/3&
$\mathbf{393822}$,
$\mathbf{494208}$
\\
&10/3&
$\mathbf{3675672}$,
$\mathbf{6675669}$,
$\mathbf{17918901}$,
$\mathbf{37459422}$
\\
&13/3&
$\mathbf{41442192}$,
$\mathbf{64849356}$,
$\mathbf{194548068'}$,
$\mathbf{310240854}$,
$\mathbf{389321856}$
\\
&16/3&
$\mathbf{5895396}$
\\
\hline
$\sigma_2$ &14/15&
$\mathbf{351'}$
\\
&29/15&
$\mathbf{54054}$
\\
&44/15&
$\mathbf{853281}$,
$\mathbf{4582656}$,
$\mathbf{6243237}$,
$\mathbf{7601958}$
\\
&59/15&
$\mathbf{30718116}$,
$\mathbf{93459366}$,
$\mathbf{159629184}$,
$\mathbf{194548068}$,
$\mathbf{219490128}$
\\
&74/15&
$\mathbf{32424678}$,
$\mathbf{70744752}$
\end{tabular}
\end{center}
\caption{Integrable highest weight representations of affine $E_6$ (labeled by the corresponding $E_6$ irrep $\mathbf{R}$) that can couple to Potts model primary $\alpha$ in the elliptic genus. The corresponding conformal weight is denoted by $h_{\mathbf{R}}$. The content for $\psi_2$ (resp. $\sigma_1$) is given by the complex conjugates of the representations appearing for $\psi_1$ (resp. $\sigma_2$).}
\label{tab:alphareps}
\end{table}
\begin{table}
\begin{center}
\begin{tabular}{c|r}
Orbit & Orbit elements \\
\hline\hline
$\mathcal{O}_{3}^{(1)}$&
$(\mathbf{1},\mathbf{5895396},\mathbf{\overline{5895396}})$
\\
$\mathcal{O}_{3}^{(2)}$&
$(\mathbf{78},\mathbf{32424678},\mathbf{\overline{32424678}})$
\\
$\mathcal{O}_{3}^{(3)}$&
$(\mathbf{34749},\mathbf{30718116},\mathbf{\overline{30718116}})$
\\
$\mathcal{O}_{3}^{(4)}$&
$(\mathbf{85293},\mathbf{3675672},\mathbf{\overline{3675672}})$
\\
$\mathcal{O}_{3}^{(5)}$&
$(\mathbf{537966},\mathbf{41442192},\mathbf{\overline{41442192}})$
\\
$\mathcal{O}_{3}^{(6)}$&
$(\mathbf{2453814},\mathbf{219490128},\mathbf{\overline{219490128}})$
\\
$\mathcal{O}_{3}^{(7)}$&
$(\mathbf{64205141},\mathbf{7601958},\mathbf{\overline{7601958}})$
\\
$\mathcal{O}_{3}^{(8)}$&
$(\mathbf{12514788},\mathbf{37459422},\mathbf{\overline{37459422}})$
\\
$\mathcal{O}_{3}^{(9)}$&
$(\mathbf{226459233},\mathbf{17918901},\mathbf{\overline{17918901}})$
\\
$\mathcal{O}_{3}^{(10)}$&
$(\mathbf{29422393},\mathbf{93459366},\mathbf{\overline{93459366}})$
\\
\hline
$\mathcal{O}_{6}^{(1)}$&
$(
\mathbf{442442}+\mathbf{\overline{442442}},
\mathbf{351}'+\mathbf{70744752},
\mathbf{\overline{351'}}+\mathbf{\overline{70744752}})$
\\
$\mathcal{O}_{6}^{(2)}$&
$(
\mathbf{15359058}+\mathbf{\overline{15359058}},
\mathbf{7371}+\mathbf{64849356},
\mathbf{\overline{7371}}+\mathbf{\overline{64849356}})$
\\
$\mathcal{O}_{6}^{(3)}$&
$(\mathbf{314269956}+\mathbf{\overline{314269956}},\mathbf{54054}+\mathbf{853281},
\mathbf{\overline{54054}}+\mathbf{\overline{853281}})$
\\
$\mathcal{O}_{6}^{(4)}$&
$(
\mathbf{89791416}+\mathbf{\overline{89791416}},
\mathbf{393822}+\mathbf{310240854},
\mathbf{\overline{393822}}+\mathbf{\overline{310240854}})$
\\
$\mathcal{O}_{6}^{(5)}$&
$(\mathbf{3309696}+\mathbf{\overline{3309696}},
\mathbf{494208}+\mathbf{389321856},
\mathbf{\overline{494208}}+\mathbf{\overline{389321856}})$
\\
$\mathcal{O}_{6}^{(6)}$&
$(\mathbf{1559376}+\mathbf{\overline{1559376}},
\mathbf{6675669}+\mathbf{194548068'},
\mathbf{\overline{6675669}}+\mathbf{\overline{194548068'}})$
\\
$\mathcal{O}_{6}^{(7)}$&
$(\mathbf{53557504}+\mathbf{\overline{53557504}},
\mathbf{4582656}+\mathbf{159629184},
\mathbf{\overline{4582656}}+\mathbf{\overline{159629184}})$
\\
$\mathcal{O}_{6}^{(8)}$&
$(\mathbf{47783736}+\mathbf{\overline{47783736}},
\mathbf{6243237}+\mathbf{194548068},
\mathbf{\overline{6243237}}+\mathbf{\overline{194548068}})$
\end{tabular}
\end{center}
\caption{$S_3$ orbits of integrable highest weight representations of affine $E_6$ that can appear in the elliptic genus. The length of the orbit is indicated by the subscript. Orbits are arranged into vectors $(\mathbf{R}_0,\mathbf{R}_1,\mathbf{R}_2)$ with component $\mathbf{R}_j$ labeled by an element $ j\in\mathbb{Z}_3$; charge conjugation acts by exchanging $\mathbf{R}_1$ and $\mathbf{R}_2$.}
\label{tab:orbits}
\end{table}

There are a total of 78 integrable highest weight representations of $E_{6,8}$ which satisfy $h^{Potts}_\alpha+h^{E_{6,8}}_{\lambda}\in\mathbb{Z}$ and therefore can couple to the Potts model primary $\alpha$; these are displayed in Table \ref{tab:alphareps}. Here we make use of the fact, justified in Appendix \ref{app:consmod} below, that primaries of given $\mathbb{Z}_3$ charge within the Potts model pair up with $E_{6,8}$ operators with opposite charge under the $\mathbb{Z}_3$ center of $E_6$. In Table \ref{tab:orbits}, on the other hand we list the 18 orbits into which these representations organize under the $S_3$ diagram automorphism group of affine $E_6$. 

Note that the range of conformal weights of $E_{6,8}$ primaries is bounded:
\begin{equation}
0 \leq h_\lambda \leq \frac{16}{3}.
\end{equation}
In particular, this allows us to determine the decomposition of the elliptic genus in terms of affine characters by examining only a limited set of terms in its $q$-expansion. In the main text we will need to know the spectrum of possible $E_{6,8}$ integrable highest weight representations that can appear at low values of the conformal weight, which is quite constrained:
\begin{align}
\widetilde{\chi}^{E_{6,8}}_1
&\simeq
n_{1,\mathbf{1}}q^{-\frac{13}{10}}(\mathbf{1}+\mathbf{78}q)+\mathcal{O}(q^{\frac{7}{10}});
\\
\nonumber
\widetilde{\chi}^{E_{6,8}}_\epsilon 
&\simeq
n_{\epsilon,\mathbf{78}}\mathbf{78}q^{-\frac{7}{10}}
+
(\mathbf{34749}n_{\epsilon,\mathbf{34749}}+
(\mathbf{1}+
\mathbf{78}+
\mathbf{650}+
\mathbf{2430}+
\mathbf{2925}
)n_{\epsilon,\mathbf{78}})q^{\frac{3}{10}}\\
&+
\mathcal{O}(q^{\frac{13}{10}});\\
\widetilde{\chi}^{E_{6,8}}_{\psi_1}
&\simeq
n_{\psi_1,\mathbf{7371}}\mathbf{7371}q^{\frac{1}{30}}+\mathcal{O}(q^{\frac{31}{30}});\\
\widetilde{\chi}^{E_{6,8}}_{\psi_2}
&\simeq
n_{\psi_2,\mathbf{\overline{7371}}}\mathbf{\overline{7371}}q^{\frac{1}{30}}+\mathcal{O}(q^{\frac{31}{30}});
\\
\nonumber
\widetilde{\chi}^{E_{6,8}}_{\sigma_1}
&\simeq
n_{\sigma_1,\mathbf{\overline{351}'}}\mathbf{\overline{351}'}q^{-\frac{11}{30}}
+
(\mathbf{\overline{54054}} n_{\epsilon,\mathbf{\overline{54054}}}
+
(
\mathbf{\overline{351}}+
\mathbf{\overline{351}'}+
\mathbf{\overline{7371}}+
\mathbf{\overline{19305}}) n_{\epsilon,\mathbf{351}'})q^{\frac{19}{30}}
\\
&+
\mathcal{O}(q^{\frac{49}{30}});
\\
\nonumber
\widetilde{\chi}^{E_{6,8}}_{\sigma_2}
&\simeq 
n_{\sigma_2,\mathbf{{351'}}}\mathbf{{351}'}q^{-\frac{11}{30}}
+
(\mathbf{54054} n_{\epsilon,\mathbf{54054}}+
(
\mathbf{351}+
\mathbf{351'}+
\mathbf{7371}+
\mathbf{19305}) n_{\epsilon,\mathbf{351}'})q^{\frac{19}{30}}
\\
&+
\mathcal{O}(q^{\frac{49}{30}}).
\end{align}

We have computed the $q$-expansion of integrable highest weight representation of $E_{6,8}$ to powers sufficient to determine the contribution up to the sixth excited level above the vacuum for $\widetilde{\chi}^{E_{6,8}}_1$ and up to the fifth excited level above the ground state for the other characters. Computations were performed using the SAGE \footnote{ \texttt{https://www.sagemath.org}} platform.

\subsection{Constraints from modularity}
\label{app:consmod}
Modular invariance of the elliptic genus 
\begin{equation}
\mathbb{E} = \sum_\alpha \chi^{Potts}_\alpha \widetilde{\chi}^{E_{6,8}}_\alpha.
\label{eq:egdec}
\end{equation}
imposes the following modular transformation on the combinations of $E_{6,8}$ characters that contribute:
\begin{equation}
\widetilde{\chi}^{E_{6,8}}_j(\vec{m}_{E_6}/\tau,-1/\tau)
=
\mathcal{S}^{E_6}
\cdot
\widetilde{\chi}^{E_{6,8}}(\vec{m}_{E_6},\tau),
\end{equation}
where 
\begin{equation}
\mathcal{S}^{E_6}
=
(\mathcal{S}^{Potts})^{-1}
=\mathcal{C}^{Potts}\mathcal{S}^{Potts},
\end{equation}
while the $T$ transformation is given by
\begin{equation}
\widetilde{\chi}^{E_{6,8}}_j(\vec{m}_{E_6},\tau+1)
=
e(-1/3)(\mathcal{T}^{Potts})^{-1}
\cdot
\widetilde{\chi}^{E_{6,8}}(\vec{m}_{E_6},\tau).
\end{equation}
Our aim here is to write down the most general possible vvmf's transforming under $SL(2,\mathbb{Z})$ with $S$ and $T$ transformations given by
\begin{equation}
\mathcal{S}
=
\mathcal{S}^{Potts},
\qquad
\mathcal{T}
=
e(-\tfrac{1}{3})(\mathcal{T}^{Potts})^{-1}
=
e(-\frac{3}{10})
\begin{pmatrix}
1 & & &\\
& e(\tfrac{1}{5}) & &\\
& & e(\tfrac{1}{3}) &\\
& & & e(\tfrac{14}{15})
\end{pmatrix}.
\end{equation}
Luckily there is a procedure to determine the entire space of vvmf's with this properties given a single element of this space, due to Bantay and Gannon \cite{Bantay:2007zz} and further discussed in \cite{Gannon:2013jua,Cheng:2020srs}. It is easy to check that such a representative $v_\star(\tau)$ can be constructed in terms of the Potts model characters as follows:
\begin{equation}
v_\star(\tau)
=
\frac{E_4(\tau)}{\eta(\tau)^8}
\begin{pmatrix}
\chi^{Potts}_\sigma\\
-\chi^{Potts}_\psi\\
-\chi^{Potts}_\epsilon\\
\chi^{Potts}_1
\end{pmatrix}
.
\end{equation}
The space of weight zero vvmf's satisfying these properties is a rank-four free module whose elements can be expressed in the form $p(j(\tau))_n v_n$, where $v_{1,\dots,4}$ is a four-dimensional basis of the module and $p(j(\tau))_n$ are polynomials in the $j$-invariant. A basis of the module can be obtained following the methods of \cite{Bantay:2007zz} by acting on $v_*$ by a suitable set of operators constructed in terms of Serre derivatives and $j$-invariants. We refer the reader to the existing literature \cite{Bantay:2007zz,Gannon:2013jua,Cheng:2020srs} for more details on the approach. The upshot is that we find a basis with the following $q$-expansion:\footnote{ While we only report the first few components here, we have determined the expansion up to $O(q^{10})$.}
\begin{align}
v_1(\tau)
&=&
\begin{pmatrix}
q^{-\frac{3}{10}}+60 q^{\frac{7}{10}}+810 q^{\frac{17}{10}}+5922 q^{\frac{27}{10}}\dots \\
26 q^{\frac{3}{10}}+507 q^{\frac{13}{10}}+4536 q^{\frac{23}{10}}\dots\\
6 q^{\frac{1}{30}}+330 q^{\frac{31}{30}}+3252 q^{\frac{61}{30}}+\dots \\
156q^{\frac{19}{30}}+2262 q^{\frac{49}{30}}+16998 q^{\frac{79}{30}}\dots 
\end{pmatrix};
\label{eq:v1}
\\
v_2(\tau)
&=&
\begin{pmatrix}
2640 q^{\frac{7}{10}}+186186 q^{\frac{17}{10}}+4694382 q^{\frac{27}{10}}+\dots \\
q^{-\frac{7}{10}}-126 q^{\frac{3}{10}}-25158 q^{\frac{13}{10}}-862540 q^{\frac{23}{10}}\dots\\
42 q^{\frac{1}{30}}+26124 q^{\frac{31}{30}}+1179444 q^{\frac{61}{30}}+\dots \\
-2310 q^{\frac{19}{30}}-180510 q^{\frac{49}{30}}-4769688 q^{\frac{79}{30}}\dots 
\end{pmatrix};
\label{eq:v2}
\\
v_3(\tau)
&=&
\begin{pmatrix}
8208 q^{\frac{7}{10}}+1383561 q^{\frac{17}{10}}+65028906 q^{\frac{27}{10}}+\dots \\
702 q^{\frac{3}{10}}+360828 q^{\frac{13}{10}}+24796044 q^{\frac{23}{10}}\dots\\
q^{-\frac{29}{30}}-18 q^{\frac{1}{30}}-57398 q^{\frac{31}{30}}-5477604 q^{\frac{61}{30}}+\dots \\
-8645 q^{\frac{19}{30}}-1675610 q^{\frac{49}{30}}-83293530 q^{\frac{79}{30}}\dots 
\end{pmatrix};
\label{eq:v3}
\\
v_4(\tau)
&=&
\begin{pmatrix}
189 q^{\frac{7}{10}}+3564 q^{\frac{17}{10}}+33453 q^{\frac{27}{10}}+\dots \\
-27 q^{\frac{3}{10}}-756 q^{\frac{13}{10}}-8910 q^{\frac{23}{10}}\dots\\
-7 q^{\frac{1}{30}}-580 q^{\frac{31}{30}}-7874 q^{\frac{61}{30}}+\dots \\
q^{-\frac{11}{30}}+ 92 q^{\frac{19}{30}}+1863 q^{\frac{49}{30}}+18004 q^{\frac{79}{30}}\dots 
\end{pmatrix}.
\label{eq:v4}
\end{align}

The modular properties lead us to the following expectation for the vector of $E_{6,8}$ character combinations that appears in Equation \eqref{eq:egdec}, in the unrefined limit $\vec{m}_{E_{6}}\to 0$:
\begin{align}
\nonumber
&
\widetilde{\chi}^{E_{6,8}}(0,\tau)
=
(\alpha_0 (j-744) + \alpha_1)v_1(\tau)
+
\alpha_2 v_2(\tau)
+
\alpha_3 v_3(\tau)
+
\alpha_4 v_4(\tau)\\
&
=
\begin{pmatrix}
\alpha_0 q^{-\frac{13}{10}}+(60\alpha_0+\alpha_1)q^{-\frac{3}{10}}+\mathcal{O}(q^{\frac{7}{10}})\\
(26\alpha_0+\alpha_2) q^{-\frac{7}{10}}+
(507\alpha_0+26\alpha_1-126\alpha_2+702\alpha_3-27\alpha_4)
q^{\frac{3}{10}}+\mathcal{O}(q^{\frac{13}{10}})\\
(6\alpha_0+\alpha_3) q^{-\frac{29}{30}}+\mathcal{O}(q^{\frac{1}{30}})\\
(156\alpha_0+\alpha_4) q^{-\frac{11}{30}}+
(2262\alpha_0+156\alpha_1-2310\alpha_2+8645\alpha_3+92\alpha_4)
q^{\frac{19}{30}}+\mathcal{O}(q^{\frac{49}{30}})\\
\end{pmatrix}
.
\label{eq:modans}
\end{align}

\bibliographystyle{JHEP}
\bibliography{refs} 

\providecommand{\href}[2]{#2}\begingroup\raggedright\begin{thebibliography}{100}

\bibitem{KimShiuVafa2019}
H.-C. Kim, G.~Shiu and C.~Vafa, \emph{Branes and the swampland}, {\emph{arXiv
  preprint} (2019) }, [\href{https://arxiv.org/abs/1905.08261}{{\ttfamily
  1905.08261}}].

\bibitem{Lee:2019skh}
S.-J. Lee and T.~Weigand, \emph{{Swampland Bounds on the Abelian Gauge
  Sector}}, \href{http://dx.doi.org/10.1103/PhysRevD.100.026015}{\emph{Phys.
  Rev. D} {\bfseries 100} (2019) 026015},
  [\href{https://arxiv.org/abs/1905.13213}{{\ttfamily 1905.13213}}].

\bibitem{Kim:2019ths}
H.-C. Kim, H.-C. Tarazi and C.~Vafa, \emph{{Four-dimensional
  $\mathbf{\mathcal{N}=4}$ SYM theory and the swampland}},
  \href{http://dx.doi.org/10.1103/PhysRevD.102.026003}{\emph{Phys. Rev. D}
  {\bfseries 102} (2020) 026003},
  [\href{https://arxiv.org/abs/1912.06144}{{\ttfamily 1912.06144}}].

\bibitem{Katz:2020ewz}
S.~Katz, H.-C. Kim, H.-C. Tarazi and C.~Vafa, \emph{{Swampland Constraints on
  5d $\mathcal{N}=1$ Supergravity}},
  \href{http://dx.doi.org/10.1007/JHEP07(2020)080}{\emph{JHEP} {\bfseries 07}
  (2020) 080}, [\href{https://arxiv.org/abs/2004.14401}{{\ttfamily
  2004.14401}}].

\bibitem{Angelantonj:2020pyr}
C.~Angelantonj, Q.~Bonnefoy, C.~Condeescu and E.~Dudas, \emph{{String Defects,
  Supersymmetry and the Swampland}},
  \href{http://dx.doi.org/10.1007/JHEP11(2020)125}{\emph{JHEP} {\bfseries 11}
  (2020) 125}, [\href{https://arxiv.org/abs/2007.12722}{{\ttfamily
  2007.12722}}].

\bibitem{Hamada:2021bbz}
Y.~Hamada and C.~Vafa, \emph{{8d supergravity, reconstruction of internal
  geometry and the Swampland}},
  \href{http://dx.doi.org/10.1007/JHEP06(2021)178}{\emph{JHEP} {\bfseries 06}
  (2021) 178}, [\href{https://arxiv.org/abs/2104.05724}{{\ttfamily
  2104.05724}}].

\bibitem{Tarazi:2021duw}
H.-C. Tarazi and C.~Vafa, \emph{{On The Finiteness of 6d Supergravity
  Landscape}},  \href{https://arxiv.org/abs/2106.10839}{{\ttfamily
  2106.10839}}.

\bibitem{Cheng:2021zjh}
P.~Cheng, R.~Minasian and S.~Theisen, \emph{{Anomalies as obstructions: from
  dimensional lifts to swampland}},
  \href{http://dx.doi.org/10.1007/JHEP01(2022)068}{\emph{JHEP} {\bfseries 01}
  (2022) 068}, [\href{https://arxiv.org/abs/2106.14912}{{\ttfamily
  2106.14912}}].

\bibitem{Cvetic:2021vsw}
M.~Cvetic, L.~Lin and A.~P. Turner, \emph{{Flavor symmetries and automatic
  enhancement in the 6D supergravity swampland}},
  \href{http://dx.doi.org/10.1103/PhysRevD.105.046005}{\emph{Phys. Rev. D}
  {\bfseries 105} (2022) 046005},
  [\href{https://arxiv.org/abs/2110.00008}{{\ttfamily 2110.00008}}].

\bibitem{Bedroya:2021fbu}
A.~Bedroya, Y.~Hamada, M.~Montero and C.~Vafa, \emph{{Compactness of brane
  moduli and the String Lamppost Principle in d \ensuremath{>} 6}},
  \href{http://dx.doi.org/10.1007/JHEP02(2022)082}{\emph{JHEP} {\bfseries 02}
  (2022) 082}, [\href{https://arxiv.org/abs/2110.10157}{{\ttfamily
  2110.10157}}].

\bibitem{Long:2021jlv}
C.~Long, M.~Montero, C.~Vafa and I.~Valenzuela, \emph{{The desert and the
  swampland}}, \href{http://dx.doi.org/10.1007/JHEP03(2023)109}{\emph{JHEP}
  {\bfseries 03} (2023) 109},
  [\href{https://arxiv.org/abs/2112.11467}{{\ttfamily 2112.11467}}].

\bibitem{Dierigl:2022zll}
M.~Dierigl, P.-K. Oehlmann and T.~Schimannek, \emph{{The discrete Green-Schwarz
  mechanism in 6D F-theory and elliptic genera of non-critical strings}},
  \href{http://dx.doi.org/10.1007/JHEP03(2023)090}{\emph{JHEP} {\bfseries 03}
  (2023) 090}, [\href{https://arxiv.org/abs/2212.04503}{{\ttfamily
  2212.04503}}].

\bibitem{Hayashi:2023hqa}
H.~Hayashi, H.-C. Kim and M.~Kim, \emph{{Spectra of BPS strings in 6d
  supergravity and the Swampland}},
  \href{http://dx.doi.org/10.1007/JHEP03(2025)123}{\emph{JHEP} {\bfseries 03}
  (2025) 123}, [\href{https://arxiv.org/abs/2310.12219}{{\ttfamily
  2310.12219}}].

\bibitem{Kim:2024tdh}
H.-C. Kim and C.~Vafa, \emph{{Exploring new constraints on K\"ahler moduli
  space of 6d $ \mathcal{N} $ = 1 supergravity}},
  \href{http://dx.doi.org/10.1007/JHEP10(2024)217}{\emph{JHEP} {\bfseries 10}
  (2024) 217}, [\href{https://arxiv.org/abs/2406.06704}{{\ttfamily
  2406.06704}}].

\bibitem{KimVafaXu2024}
H.-C. Kim, C.~Vafa and K.~Xu, \emph{Finite landscape of 6d $\mathcal{N} =
  (1,0)$ supergravity}, {\emph{arXiv preprint} (2024) },
  [\href{https://arxiv.org/abs/2411.19155}{{\ttfamily 2411.19155}}].

\bibitem{Grimm:2018ohb}
T.~W. Grimm, E.~Palti and I.~Valenzuela, \emph{{Infinite Distances in Field
  Space and Massless Towers of States}},
  \href{http://dx.doi.org/10.1007/JHEP08(2018)143}{\emph{JHEP} {\bfseries 08}
  (2018) 143}, [\href{https://arxiv.org/abs/1802.08264}{{\ttfamily
  1802.08264}}].

\bibitem{Blumenhagen:2018nts}
R.~Blumenhagen, D.~Kl\"awer, L.~Schlechter and F.~Wolf, \emph{{The Refined
  Swampland Distance Conjecture in Calabi-Yau Moduli Spaces}},
  \href{http://dx.doi.org/10.1007/JHEP06(2018)052}{\emph{JHEP} {\bfseries 06}
  (2018) 052}, [\href{https://arxiv.org/abs/1803.04989}{{\ttfamily
  1803.04989}}].

\bibitem{Lee:2018urn}
S.-J. Lee, W.~Lerche and T.~Weigand, \emph{{Tensionless Strings and the Weak
  Gravity Conjecture}},
  \href{http://dx.doi.org/10.1007/JHEP10(2018)164}{\emph{JHEP} {\bfseries 10}
  (2018) 164}, [\href{https://arxiv.org/abs/1808.05958}{{\ttfamily
  1808.05958}}].

\bibitem{Grimm:2018cpv}
T.~W. Grimm, C.~Li and E.~Palti, \emph{{Infinite Distance Networks in Field
  Space and Charge Orbits}},
  \href{http://dx.doi.org/10.1007/JHEP03(2019)016}{\emph{JHEP} {\bfseries 03}
  (2019) 016}, [\href{https://arxiv.org/abs/1811.02571}{{\ttfamily
  1811.02571}}].

\bibitem{Joshi:2019nzi}
A.~Joshi and A.~Klemm, \emph{{Swampland Distance Conjecture for One-Parameter
  Calabi-Yau Threefolds}},
  \href{http://dx.doi.org/10.1007/JHEP08(2019)086}{\emph{JHEP} {\bfseries 08}
  (2019) 086}, [\href{https://arxiv.org/abs/1903.00596}{{\ttfamily
  1903.00596}}].

\bibitem{Font:2019cxq}
A.~Font, A.~Herr\'aez and L.~E. Ib\'a\~nez, \emph{{The Swampland Distance
  Conjecture and Towers of Tensionless Branes}},
  \href{http://dx.doi.org/10.1007/JHEP08(2019)044}{\emph{JHEP} {\bfseries 08}
  (2019) 044}, [\href{https://arxiv.org/abs/1904.05379}{{\ttfamily
  1904.05379}}].

\bibitem{Lee:2019xtm}
S.-J. Lee, W.~Lerche and T.~Weigand, \emph{{Emergent strings, duality and weak
  coupling limits for two-form fields}},
  \href{http://dx.doi.org/10.1007/JHEP02(2022)096}{\emph{JHEP} {\bfseries 02}
  (2022) 096}, [\href{https://arxiv.org/abs/1904.06344}{{\ttfamily
  1904.06344}}].

\bibitem{Erkinger:2019umg}
D.~Erkinger and J.~Knapp, \emph{{Refined swampland distance conjecture and
  exotic hybrid Calabi-Yaus}},
  \href{http://dx.doi.org/10.1007/JHEP07(2019)029}{\emph{JHEP} {\bfseries 07}
  (2019) 029}, [\href{https://arxiv.org/abs/1905.05225}{{\ttfamily
  1905.05225}}].

\bibitem{Lee:2019wij}
S.-J. Lee, W.~Lerche and T.~Weigand, \emph{{Emergent strings from infinite
  distance limits}},
  \href{http://dx.doi.org/10.1007/JHEP02(2022)190}{\emph{JHEP} {\bfseries 02}
  (2022) 190}, [\href{https://arxiv.org/abs/1910.01135}{{\ttfamily
  1910.01135}}].

\bibitem{Grimm:2019ixq}
T.~W. Grimm, C.~Li and I.~Valenzuela, \emph{{Asymptotic Flux Compactifications
  and the Swampland}},
  \href{http://dx.doi.org/10.1007/JHEP06(2020)009}{\emph{JHEP} {\bfseries 06}
  (2020) 009}, [\href{https://arxiv.org/abs/1910.09549}{{\ttfamily
  1910.09549}}].

\bibitem{Baume:2019sry}
F.~Baume, F.~Marchesano and M.~Wiesner, \emph{{Instanton Corrections and
  Emergent Strings}},
  \href{http://dx.doi.org/10.1007/JHEP04(2020)174}{\emph{JHEP} {\bfseries 04}
  (2020) 174}, [\href{https://arxiv.org/abs/1912.02218}{{\ttfamily
  1912.02218}}].

\bibitem{Bastian:2020egp}
B.~Bastian, T.~W. Grimm and D.~van~de Heisteeg, \emph{{Weak gravity bounds in
  asymptotic string compactifications}},
  \href{http://dx.doi.org/10.1007/JHEP06(2021)162}{\emph{JHEP} {\bfseries 06}
  (2021) 162}, [\href{https://arxiv.org/abs/2011.08854}{{\ttfamily
  2011.08854}}].

\bibitem{Bastian:2021eom}
B.~Bastian, T.~W. Grimm and D.~van~de Heisteeg, \emph{{Modeling General
  Asymptotic Calabi-Yau Periods}},
  \href{https://arxiv.org/abs/2105.02232}{{\ttfamily 2105.02232}}.

\bibitem{Palti:2021ubp}
E.~Palti, \emph{{Stability of BPS states and weak coupling limits}},
  \href{http://dx.doi.org/10.1007/JHEP08(2021)091}{\emph{JHEP} {\bfseries 08}
  (2021) 091}, [\href{https://arxiv.org/abs/2107.01539}{{\ttfamily
  2107.01539}}].

\bibitem{Klawer:2021ltm}
D.~Kl\"awer, \emph{{Modular curves and the refined distance conjecture}},
  \href{http://dx.doi.org/10.1007/JHEP12(2021)088}{\emph{JHEP} {\bfseries 12}
  (2021) 088}, [\href{https://arxiv.org/abs/2108.00021}{{\ttfamily
  2108.00021}}].

\bibitem{Alvarez-Garcia:2021mzv}
R.~\'Alvarez-Garc\'\i{}a and L.~Schlechter, \emph{{Analytic periods via twisted
  symmetric squares}},
  \href{http://dx.doi.org/10.1007/JHEP07(2022)024}{\emph{JHEP} {\bfseries 07}
  (2022) 024}, [\href{https://arxiv.org/abs/2110.02962}{{\ttfamily
  2110.02962}}].

\bibitem{Alvarez-Garcia:2021pxo}
R.~\'Alvarez-Garc\'\i{}a, D.~Kl\"awer and T.~Weigand, \emph{{Membrane limits in
  quantum gravity}},
  \href{http://dx.doi.org/10.1103/PhysRevD.105.066024}{\emph{Phys. Rev. D}
  {\bfseries 105} (2022) 066024},
  [\href{https://arxiv.org/abs/2112.09136}{{\ttfamily 2112.09136}}].

\bibitem{Grimm:2022sbl}
T.~W. Grimm, S.~Lanza and C.~Li, \emph{{Tameness, Strings, and the Distance
  Conjecture}}, \href{http://dx.doi.org/10.1007/JHEP09(2022)149}{\emph{JHEP}
  {\bfseries 09} (2022) 149},
  [\href{https://arxiv.org/abs/2206.00697}{{\ttfamily 2206.00697}}].

\bibitem{Rudelius:2023odg}
T.~Rudelius, \emph{{Gopakumar-Vafa invariants and the Emergent String
  Conjecture}}, \href{http://dx.doi.org/10.1007/JHEP03(2024)061}{\emph{JHEP}
  {\bfseries 03} (2024) 061},
  [\href{https://arxiv.org/abs/2309.10024}{{\ttfamily 2309.10024}}].

\bibitem{Alvarez-Garcia:2023gdd}
R.~\'Alvarez-Garc\'\i{}a, S.-J. Lee and T.~Weigand, \emph{{Non-minimal elliptic
  threefolds at infinite distance. Part I. Log Calabi-Yau resolutions}},
  \href{http://dx.doi.org/10.1007/JHEP08(2024)240}{\emph{JHEP} {\bfseries 08}
  (2024) 240}, [\href{https://arxiv.org/abs/2310.07761}{{\ttfamily
  2310.07761}}].

\bibitem{Alvarez-Garcia:2023qqj}
R.~\'Alvarez-Garc\'\i{}a, S.-J. Lee and T.~Weigand, \emph{{Non-minimal elliptic
  threefolds at infinite distance II: asymptotic physics}},
  \href{http://dx.doi.org/10.1007/JHEP01(2025)058}{\emph{JHEP} {\bfseries 01}
  (2025) 058}, [\href{https://arxiv.org/abs/2312.11611}{{\ttfamily
  2312.11611}}].

\bibitem{Aoufia:2025ppe}
C.~Aoufia, A.~Castellano and L.~Ib\'a\~nez, \emph{{Laplacians in Various
  Dimensions and the Swampland}},
  \href{https://arxiv.org/abs/2506.03253}{{\ttfamily 2506.03253}}.

\bibitem{Kumar:2010ru}
V.~Kumar, D.~R. Morrison and W.~Taylor, \emph{{Global aspects of the space of
  6D N = 1 supergravities}},
  \href{http://dx.doi.org/10.1007/JHEP11(2010)118}{\emph{JHEP} {\bfseries 11}
  (2010) 118}, [\href{https://arxiv.org/abs/1008.1062}{{\ttfamily 1008.1062}}].

\bibitem{Kachru:1995wm}
S.~Kachru and C.~Vafa, \emph{{Exact results for N=2 compactifications of
  heterotic strings}},
  \href{http://dx.doi.org/10.1016/0550-3213(95)00307-E}{\emph{Nucgitl. Phys. B}
  {\bfseries 450} (1995) 69--89},
  [\href{https://arxiv.org/abs/hep-th/9505105}{{\ttfamily hep-th/9505105}}].

\bibitem{Gkountoumis:2023fym}
G.~Gkountoumis, C.~Hull, K.~Stemerdink and S.~Vandoren, \emph{{Freely acting
  orbifolds of type IIB string theory on T$^{5}$}},
  \href{http://dx.doi.org/10.1007/JHEP08(2023)089}{\emph{JHEP} {\bfseries 08}
  (2023) 089}, [\href{https://arxiv.org/abs/2302.09112}{{\ttfamily
  2302.09112}}].

\bibitem{Israel:2013wwa}
D.~Isra\"el and V.~Thi\'ery, \emph{{Asymmetric Gepner models in type II}},
  \href{http://dx.doi.org/10.1007/JHEP02(2014)011}{\emph{JHEP} {\bfseries 02}
  (2014) 011}, [\href{https://arxiv.org/abs/1310.4116}{{\ttfamily 1310.4116}}].

\bibitem{Dolivet:2007sz}
Y.~Dolivet, B.~Julia and C.~Kounnas, \emph{{Magic N=2 supergravities from
  hyper-free superstrings}},
  \href{http://dx.doi.org/10.1088/1126-6708/2008/02/097}{\emph{JHEP} {\bfseries
  02} (2008) 097}, [\href{https://arxiv.org/abs/0712.2867}{{\ttfamily
  0712.2867}}].

\bibitem{HamadaLoges2023}
Y.~Hamada and G.~J. Loges, \emph{Towards a complete classification of 6d
  supergravities}, {\emph{arXiv preprint} (2023) },
  [\href{https://arxiv.org/abs/2311.00868}{{\ttfamily 2311.00868}}].

\bibitem{HaghighatMurthyVafaVandoren2015}
B.~Haghighat, S.~Murthy, C.~Vafa and S.~Vandoren, \emph{F-theory, spinning
  black holes and multi-string branches},
  \href{https://arxiv.org/abs/1509.00455}{{\ttfamily 1509.00455}}.

\bibitem{BaykaraHamadaTaraziVafa2023}
Z.~K. Baykara, Y.~Hamada, H.-C. Tarazi and C.~Vafa, \emph{On the string
  landscape without hypermultiplets}, {\emph{arXiv preprint} (2023) },
  [\href{https://arxiv.org/abs/2309.15152}{{\ttfamily 2309.15152}}].

\bibitem{Baykara:2024vss}
Z.~K. Baykara, H.-C. Tarazi and C.~Vafa, \emph{{Quasicrystalline string
  landscape}}, \href{http://dx.doi.org/10.1103/PhysRevD.111.086025}{\emph{Phys.
  Rev. D} {\bfseries 111} (2025) 086025},
  [\href{https://arxiv.org/abs/2406.00129}{{\ttfamily 2406.00129}}].

\bibitem{Melnikov:2019tpl}
I.~V. Melnikov, \emph{{An Introduction to Two-Dimensional Quantum Field Theory
  with (0,2) Supersymmetry}}, vol.~951 of \emph{Lecture Notes in Physics}.
\newblock Springer, 2019,
  \href{http://dx.doi.org/10.1007/978-3-030-05085-6}{10.1007/978-3-030-05085-6}.

\bibitem{wip}
G.~Lockhart, L.~Novelli and Y.~Proto, \emph{{Work in progress}}, .

\bibitem{HuangKatzKlemm2015}
M.~xin Huang, S.~Katz and A.~Klemm, \emph{Topological string on elliptic
  calabi--yau 3-folds and the ring of jacobi forms},
  \href{https://arxiv.org/abs/1501.04891}{{\ttfamily 1501.04891}}.

\bibitem{Gepner:1987qi}
D.~Gepner, \emph{{Space-Time Supersymmetry in Compactified String Theory and
  Superconformal Models}},
  \href{http://dx.doi.org/10.1016/0550-3213(88)90397-5}{\emph{Nucl. Phys. B}
  {\bfseries 296} (1988) 757}.

\bibitem{Eguchi:1988vra}
T.~Eguchi, H.~Ooguri, A.~Taormina and S.-K. Yang, \emph{{Superconformal
  Algebras and String Compactification on Manifolds with SU(N) Holonomy}},
  \href{http://dx.doi.org/10.1016/0550-3213(89)90454-9}{\emph{Nucl. Phys. B}
  {\bfseries 315} (1989) 193--221}.

\bibitem{Nahm:1999ps}
W.~Nahm and K.~Wendland, \emph{{A Hiker's guide to K3: Aspects of N=(4,4)
  superconformal field theory with central charge c = 6}},
  \href{http://dx.doi.org/10.1007/PL00005548}{\emph{Commun. Math. Phys.}
  {\bfseries 216} (2001) 85--138},
  [\href{https://arxiv.org/abs/hep-th/9912067}{{\ttfamily hep-th/9912067}}].

\bibitem{Martinec:2001hh}
E.~J. Martinec and G.~W. Moore, \emph{{Noncommutative solitons on orbifolds}},
  \href{https://arxiv.org/abs/hep-th/0101199}{{\ttfamily hep-th/0101199}}.

\bibitem{Greene:1988ut}
B.~R. Greene, C.~Vafa and N.~P. Warner, \emph{{Calabi-Yau Manifolds and
  Renormalization Group Flows}},
  \href{http://dx.doi.org/10.1016/0550-3213(89)90471-9}{\emph{Nucl. Phys. B}
  {\bfseries 324} (1989) 371}.

\bibitem{Green:1984}
M.~B. Green and J.~H. Schwarz, \emph{Anomaly cancellation in supersymmetric
  d=10 gauge theory and superstring theory},
  \href{http://dx.doi.org/10.1016/0370-2693(84)91565-X}{\emph{Phys. Lett. B}
  {\bfseries 149} (1984) 117--122}.

\bibitem{Green:1985}
M.~B. Green, J.~H. Schwarz and P.~C. West, \emph{Anomaly free chiral theories
  in six-dimensions},
  \href{http://dx.doi.org/10.1016/0550-3213(85)90222-6}{\emph{Nucl. Phys. B}
  {\bfseries 254} (1985) 327--348}.

\bibitem{Sagnotti:1992}
A.~Sagnotti, \emph{A note on the green-schwarz mechanism in open string
  theories}, \href{http://dx.doi.org/10.1016/0370-2693(92)90682-T}{\emph{Phys.
  Lett. B} {\bfseries 294} (1992) 196--203},
  [\href{https://arxiv.org/abs/hep-th/9210127}{{\ttfamily hep-th/9210127}}].

\bibitem{MorrisonPark2012}
D.~R. Morrison and D.~S. Park, \emph{F-theory and the mordell-weil group of
  elliptically-fibered calabi-yau threefolds},
  \href{http://dx.doi.org/10.1007/JHEP10(2012)128}{\emph{Journal of High Energy
  Physics} {\bfseries 2012} (2012) 128},
  [\href{https://arxiv.org/abs/1208.2695}{{\ttfamily 1208.2695}}].

\bibitem{CompactificationFtheory1}
D.~R. Morrison and C.~Vafa, \emph{Compactifications of f-theory on calabi–yau
  threefolds -- i},
  \href{http://dx.doi.org/10.1016/0550-3213(96)00242-8}{\emph{Nucl. Phys. B}
  {\bfseries 473} (1996) 74--92},
  [\href{https://arxiv.org/abs/hep-th/9602114}{{\ttfamily hep-th/9602114}}].

\bibitem{CompactificationFtheory2}
D.~R. Morrison and C.~Vafa, \emph{Compactifications of f-theory on calabi–yau
  threefolds -- ii},
  \href{http://dx.doi.org/10.1016/0550-3213(96)00369-0}{\emph{Nucl. Phys. B}
  {\bfseries 476} (1996) 437--469},
  [\href{https://arxiv.org/abs/hep-th/9603161}{{\ttfamily hep-th/9603161}}].

\bibitem{Kumar2009}
V.~Kumar, D.~R. Morrison and W.~Taylor, \emph{Mapping 6d n = 1 supergravity to
  f-theory}, {\emph{arXiv preprint} (2009) },
  [\href{https://arxiv.org/abs/0911.3393}{{\ttfamily 0911.3393}}].

\bibitem{CVSG25}
C.~Vafa, \emph{{Geometric vs Non-Geometric Landscape}}, {\emph{Talk at Strings
  and Geometry 2025} }.

\bibitem{Narain:1986qm}
K.~S. Narain, M.~H. Sarmadi and C.~Vafa, \emph{{Asymmetric Orbifolds}},
  \href{http://dx.doi.org/10.1016/0550-3213(87)90228-8}{\emph{Nucl. Phys. B}
  {\bfseries 288} (1987) 551}.

\bibitem{Narain:1990mw}
K.~S. Narain, M.~H. Sarmadi and C.~Vafa, \emph{{Asymmetric orbifolds: Path
  integral and operator formulations}},
  \href{http://dx.doi.org/10.1016/0550-3213(91)90145-N}{\emph{Nucl. Phys. B}
  {\bfseries 356} (1991) 163--207}.

\bibitem{database}
G.~Loges, \emph{{6d-sugra-data}},
  {\emph{https://github.com/gloges/6d-sugra-data, 2023} }.

\bibitem{Klevers:2017aku}
D.~Klevers, D.~R. Morrison, N.~Raghuram and W.~Taylor, \emph{Exotic matter on
  singular divisors in f-theory},
  \href{http://dx.doi.org/10.1007/JHEP10(2017)168}{\emph{JHEP} {\bfseries 10}
  (2017) 168}, [\href{https://arxiv.org/abs/1706.08194}{{\ttfamily
  1706.08194}}].

\bibitem{Candelas1993mirrorI}
P.~Candelas, X.~de~la Ossa, A.~Font, S.~Katz and D.~R. Morrison, \emph{Mirror
  symmetry for two parameter models -- i},
  \href{https://arxiv.org/abs/hep-th/9308083}{{\ttfamily hep-th/9308083}}.

\bibitem{Candelas1994mirrorII}
P.~Candelas, A.~Font, S.~Katz and D.~R. Morrison, \emph{Mirror symmetry for two
  parameter models -- ii},
  \href{https://arxiv.org/abs/hep-th/9403187}{{\ttfamily hep-th/9403187}}.

\bibitem{Hosono1995hypersurfaces}
S.~Hosono, A.~Klemm, S.~Theisen and S.-T. Yau, \emph{Mirror symmetry, mirror
  map and applications to calabi-yau hypersurfaces},
  \href{http://dx.doi.org/10.1007/BF02100589}{\emph{Communications in
  Mathematical Physics} {\bfseries 167} (1995) 301--350},
  [\href{https://arxiv.org/abs/hep-th/9308122}{{\ttfamily hep-th/9308122}}].

\bibitem{Hosono1995complete}
S.~Hosono, A.~Klemm, S.~Theisen and S.-T. Yau, \emph{Mirror symmetry, mirror
  map and applications to complete intersection calabi–yau spaces},
  \href{http://dx.doi.org/10.1016/0550-3213(94)00440-P}{\emph{Nuclear Physics
  B} {\bfseries 433} (1995) 501--552},
  [\href{https://arxiv.org/abs/hep-th/9406055}{{\ttfamily hep-th/9406055}}].

\bibitem{Polchinski:2003bq}
J.~Polchinski, \emph{{Monopoles, duality, and string theory}},
  \href{http://dx.doi.org/10.1142/S0217751X0401866X}{\emph{Int. J. Mod. Phys.
  A} {\bfseries 19S1} (2004) 145--156},
  [\href{https://arxiv.org/abs/hep-th/0304042}{{\ttfamily hep-th/0304042}}].

\bibitem{Banks:1997zs}
T.~Banks, N.~Seiberg and E.~Silverstein, \emph{{Zero and one-dimensional probes
  with N=8 supersymmetry}},
  \href{http://dx.doi.org/10.1016/S0370-2693(97)00366-3}{\emph{Phys. Lett. B}
  {\bfseries 401} (1997) 30--37},
  [\href{https://arxiv.org/abs/hep-th/9703052}{{\ttfamily hep-th/9703052}}].

\bibitem{Harlow:2018tng}
D.~Harlow and H.~Ooguri, \emph{{Symmetries in quantum field theory and quantum
  gravity}}, \href{http://dx.doi.org/10.1007/s00220-021-04040-y}{\emph{Commun.
  Math. Phys.} {\bfseries 383} (2021) 1669--1804},
  [\href{https://arxiv.org/abs/1810.05338}{{\ttfamily 1810.05338}}].

\bibitem{Lee:2018spm}
S.-J. Lee, W.~Lerche and T.~Weigand, \emph{{A Stringy Test of the Scalar Weak
  Gravity Conjecture}},
  \href{http://dx.doi.org/10.1016/j.nuclphysb.2018.11.001}{\emph{Nucl. Phys. B}
  {\bfseries 938} (2019) 321--350},
  [\href{https://arxiv.org/abs/1810.05169}{{\ttfamily 1810.05169}}].

\bibitem{Bershadsky:1995vm}
M.~Bershadsky, A.~Johansen, V.~Sadov and C.~Vafa, \emph{{Topological reduction
  of 4-d SYM to 2-d sigma models}},
  \href{http://dx.doi.org/10.1016/0550-3213(95)00242-K}{\emph{Nucl. Phys. B}
  {\bfseries 448} (1995) 166--186},
  [\href{https://arxiv.org/abs/hep-th/9501096}{{\ttfamily hep-th/9501096}}].

\bibitem{DelZotto:2017mee}
M.~Del~Zotto, J.~Gu, M.-X. Huang, A.-K. Kashani-Poor, A.~Klemm and G.~Lockhart,
  \emph{{Topological Strings on Singular Elliptic Calabi-Yau 3-folds and
  Minimal 6d SCFTs}},
  \href{http://dx.doi.org/10.1007/JHEP03(2018)156}{\emph{JHEP} {\bfseries 03}
  (2018) 156}, [\href{https://arxiv.org/abs/1712.07017}{{\ttfamily
  1712.07017}}].

\bibitem{DelZotto:2018tcj}
M.~Del~Zotto and G.~Lockhart, \emph{{Universal Features of BPS Strings in
  Six-dimensional SCFTs}},
  \href{http://dx.doi.org/10.1007/JHEP08(2018)173}{\emph{JHEP} {\bfseries 08}
  (2018) 173}, [\href{https://arxiv.org/abs/1804.09694}{{\ttfamily
  1804.09694}}].

\bibitem{Kim:2016foj}
H.-C. Kim, S.~Kim and J.~Park, \emph{{6d strings from new chiral gauge
  theories}},  \href{https://arxiv.org/abs/1608.03919}{{\ttfamily 1608.03919}}.

\bibitem{Shimizu:2016lbw}
H.~Shimizu and Y.~Tachikawa, \emph{{Anomaly of strings of 6d $
  \mathcal{N}=\left(1,0\right) $ theories}},
  \href{http://dx.doi.org/10.1007/JHEP11(2016)165}{\emph{JHEP} {\bfseries 11}
  (2016) 165}, [\href{https://arxiv.org/abs/1608.05894}{{\ttfamily
  1608.05894}}].

\bibitem{Witten:1985xe}
E.~Witten, \emph{{GLOBAL GRAVITATIONAL ANOMALIES}},
  \href{http://dx.doi.org/10.1007/BF01212448}{\emph{Commun. Math. Phys.}
  {\bfseries 100} (1985) 197}.

\bibitem{Freed:1986zx}
D.~S. Freed, \emph{{Determinants, Torsion, and Strings}},
  \href{http://dx.doi.org/10.1007/BF01221001}{\emph{Commun. Math. Phys.}
  {\bfseries 107} (1986) 483--513}.

\bibitem{Distler:1986wm}
J.~Distler, \emph{{RESURRECTING (2,0) COMPACTIFICATIONS}},
  \href{http://dx.doi.org/10.1016/0370-2693(87)91643-1}{\emph{Phys. Lett. B}
  {\bfseries 188} (1987) 431--436}.

\bibitem{Dai:1994kq}
X.-z. Dai and D.~S. Freed, \emph{{eta invariants and determinant lines}},
  \href{http://dx.doi.org/10.1063/1.530747}{\emph{J. Math. Phys.} {\bfseries
  35} (1994) 5155--5194},
  [\href{https://arxiv.org/abs/hep-th/9405012}{{\ttfamily hep-th/9405012}}].

\bibitem{Kumar:2010am}
V.~Kumar, D.~S. Park and W.~Taylor, \emph{{6D supergravity without tensor
  multiplets}}, \href{http://dx.doi.org/10.1007/JHEP04(2011)080}{\emph{JHEP}
  {\bfseries 04} (2011) 080},
  [\href{https://arxiv.org/abs/1011.0726}{{\ttfamily 1011.0726}}].

\bibitem{Dijkgraaf:1996xw}
R.~Dijkgraaf, G.~W. Moore, E.~P. Verlinde and H.~L. Verlinde, \emph{{Elliptic
  genera of symmetric products and second quantized strings}},
  \href{http://dx.doi.org/10.1007/s002200050087}{\emph{Commun. Math. Phys.}
  {\bfseries 185} (1997) 197--209},
  [\href{https://arxiv.org/abs/hep-th/9608096}{{\ttfamily hep-th/9608096}}].

\bibitem{sjunp}
S.-J. Lee, W.~Lerche, G.~Lockhart and T.~Weigand, \emph{{Unpublished notes}}, .

\bibitem{Tong:2008qd}
D.~Tong, \emph{{Quantum Vortex Strings: A Review}},
  \href{http://dx.doi.org/10.1016/j.aop.2008.10.005}{\emph{Annals Phys.}
  {\bfseries 324} (2009) 30--52},
  [\href{https://arxiv.org/abs/0809.5060}{{\ttfamily 0809.5060}}].

\bibitem{Warner:1989yy}
N.~P. Warner, \emph{{The Supersymmetry Index and the Construction of Modular
  Invariants}}, \href{http://dx.doi.org/10.1007/BF02099882}{\emph{Commun. Math.
  Phys.} {\bfseries 130} (1990) 205}.

\bibitem{Roberts:1990tv}
P.~Roberts and H.~Terao, \emph{{Modular invariants of Kac-Moody algebras from
  selfdual lattices}},
  \href{http://dx.doi.org/10.1142/S0217751X92000983}{\emph{Int. J. Mod. Phys.
  A} {\bfseries 7} (1992) 2207--2218}.

\bibitem{Gannon:1995jm}
T.~Gannon, P.~Ruelle and M.~A. Walton, \emph{{Automorphism modular invariants
  of current algebras}},
  \href{http://dx.doi.org/10.1007/BF02103717}{\emph{Commun. Math. Phys.}
  {\bfseries 179} (1996) 121--156},
  [\href{https://arxiv.org/abs/hep-th/9503141}{{\ttfamily hep-th/9503141}}].

\bibitem{Gannon:1995uv}
T.~Gannon, \emph{{Kac-Peterson, Perron-Frobenius, and the classification of
  conformal field theories}},
  \href{https://arxiv.org/abs/q-alg/9510026}{{\ttfamily q-alg/9510026}}.

\bibitem{Gaberdiel:2010ch}
M.~R. Gaberdiel, S.~Hohenegger and R.~Volpato, \emph{{Mathieu twining
  characters for K3}},
  \href{http://dx.doi.org/10.1007/JHEP09(2010)058}{\emph{JHEP} {\bfseries 09}
  (2010) 058}, [\href{https://arxiv.org/abs/1006.0221}{{\ttfamily 1006.0221}}].

\bibitem{Eguchi:2010fg}
T.~Eguchi and K.~Hikami, \emph{{Note on twisted elliptic genus of $K3$
  surface}},
  \href{http://dx.doi.org/10.1016/j.physletb.2010.10.017}{\emph{Phys. Lett. B}
  {\bfseries 694} (2011) 446--455},
  [\href{https://arxiv.org/abs/1008.4924}{{\ttfamily 1008.4924}}].

\bibitem{Harrison:2013bya}
S.~Harrison, S.~Kachru and N.~M. Paquette, \emph{{Twining Genera of (0,4)
  Supersymmetric Sigma Models on K3}},
  \href{http://dx.doi.org/10.1007/JHEP04(2014)048}{\emph{JHEP} {\bfseries 04}
  (2014) 048}, [\href{https://arxiv.org/abs/1309.0510}{{\ttfamily 1309.0510}}].

\bibitem{Cheng:2015rby}
M.~C.~N. Cheng, F.~Ferrari, S.~M. Harrison and N.~M. Paquette,
  \emph{{Landau-Ginzburg Orbifolds and Symmetries of K3 CFTs}},
  \href{http://dx.doi.org/10.1007/JHEP01(2017)046}{\emph{JHEP} {\bfseries 01}
  (2017) 046}, [\href{https://arxiv.org/abs/1512.04942}{{\ttfamily
  1512.04942}}].

\bibitem{Schimannek:2021pau}
T.~Schimannek, \emph{{Modular curves, the Tate-Shafarevich group and
  Gopakumar-Vafa invariants with discrete charges}},
  \href{http://dx.doi.org/10.1007/JHEP02(2022)007}{\emph{JHEP} {\bfseries 02}
  (2022) 007}, [\href{https://arxiv.org/abs/2108.09311}{{\ttfamily
  2108.09311}}].

\bibitem{Gadde:2014ppa}
A.~Gadde, S.~Gukov and P.~Putrov, \emph{{Exact Solutions of 2d Supersymmetric
  Gauge Theories}},
  \href{http://dx.doi.org/10.1007/JHEP11(2019)174}{\emph{JHEP} {\bfseries 11}
  (2019) 174}, [\href{https://arxiv.org/abs/1404.5314}{{\ttfamily 1404.5314}}].

\bibitem{Feger:2019tvk}
R.~Feger, T.~W. Kephart and R.~J. Saskowski, \emph{{LieART 2.0 \textendash{} A
  Mathematica application for Lie Algebras and Representation Theory}},
  \href{http://dx.doi.org/10.1016/j.cpc.2020.107490}{\emph{Comput. Phys.
  Commun.} {\bfseries 257} (2020) 107490},
  [\href{https://arxiv.org/abs/1912.10969}{{\ttfamily 1912.10969}}].

\bibitem{Bantay:2007zz}
P.~Bantay and T.~Gannon, \emph{{Vector-valued modular functions for the modular
  group and the hypergeometric equation}},
  \href{http://dx.doi.org/10.4310/CNTP.2007.v1.n4.a2}{\emph{Commun. Num. Theor.
  Phys.} {\bfseries 1} (2007) 651--680}.

\bibitem{Gannon:2013jua}
T.~Gannon, \emph{{The theory of vector-modular forms for the modular group}},
  \href{http://dx.doi.org/10.1007/978-3-662-43831-2_9}{\emph{Contrib. Math.
  Comput. Sci.} {\bfseries 8} (2014) 247--286},
  [\href{https://arxiv.org/abs/1310.4458}{{\ttfamily 1310.4458}}].

\bibitem{Cheng:2020srs}
M.~C.~N. Cheng, T.~Gannon and G.~Lockhart, \emph{{Modular Exercises for
  Four-Point Blocks - I}},
  \href{http://dx.doi.org/10.3842/SIGMA.2025.013}{\emph{SIGMA} {\bfseries 21}
  (2025) 013}, [\href{https://arxiv.org/abs/2002.11125}{{\ttfamily
  2002.11125}}].

\end{thebibliography}\endgroup

\end{document}